\documentclass[aps,reprint]{revtex4-2}
\usepackage{graphicx}
\usepackage{amsmath}
\usepackage{cases}
\usepackage{amsfonts}
\draft 

\begin{document}


\title{Topological edge states and disorder robustness in one-dimensional off-diagonal mosaic lattices}

\author{Ba Phi Nguyen}
\affiliation{Department of Basic Sciences, Mientrung University of Civil Engineering, Tuy Hoa 620000, Vietnam}
\affiliation{Mathematics and Physics Research Group, Mientrung University of Civil Engineering, Tuy Hoa 620000, Vietnam}
\author{Kihong Kim}
\email{khkim@ajou.ac.kr}
\affiliation{Department of Physics, Ajou University, Suwon 16499, South Korea}
\affiliation{School of Physics, Korea Institute for Advanced Study, Seoul 02455, South Korea}
\date{\today}
\begin{abstract}
We investigate topological edge states in one-dimensional off-diagonal mosaic lattices, where nearest-neighbor hopping amplitudes are modulated periodically with period
$\kappa>1$. Analytically, we demonstrate that discrete edge states emerge at energy levels
$E=\epsilon+2t\cos(\pi i/\kappa)$ ($i=1,\cdots,\kappa-1$), extending the Su–Schrieffer–Heeger model to multi-band systems. Numerical simulations show that these edge states are robustly localized and display characteristic nodal structures, with their existence being strongly dictated by the specific edge arrangement of long and short bonds. We further examine their stability under off-diagonal disorder, where the hopping amplitudes $\beta$ fluctuate randomly at intervals of $\kappa$. Using the inverse participation ratio as a localization measure, we show that these topological edge states remain robust over a broad range of disorder strengths. In contrast, additional $\beta$-dependent edge states that appear for
$\kappa \ge 4$
are fragile and vanish even under relatively weak disorder. These findings highlight a rich interplay between topology, periodic modulation, and disorder, offering insights for engineering multi-gap topological phases and their realization in synthetic quantum and photonic systems.
\end{abstract}

\maketitle

\section{\label{sec:level1} Introduction}

Topological insulators are quantum materials characterized by insulating bulk states and robust conducting edge or surface states \cite{Shen, Bern}. These unique properties have spurred significant interest in applications ranging from electronics and spintronics to quantum computing \cite{Moo, Has, Qi}. While early studies focused primarily on two-dimensional (2D) systems \cite{Kan1, Kan2}, recent advances in experimental techniques and the conceptual simplicity of lower-dimensional models have shifted attention toward one-dimensional (1D) systems \cite{Lan, Guo}.

To explore topological phenomena in one dimension, various lattice models have been proposed \cite{Kan1, SSH1, Hal, Asb}, incorporating both diagonal (on-site potential) and off-diagonal (hopping amplitude) modulations. Among these, the Su–Schrieffer–Heeger (SSH) model stands out as a paradigmatic example of 1D topological phases \cite{SSH1}. Originally developed to describe polyacetylene, the SSH model treats spinless fermions on a bipartite chain with alternating strong and weak bonds. The competition between the two dimerization patterns gives rise to distinct insulating phases: a trivial phase, with only extended bulk states and no boundary modes, and a nontrivial phase, characterized by topological invariants such as the Zak phase \cite{Zak} and the emergence of zero-energy edge states under open boundaries, reflecting bulk–boundary correspondence \cite{Has, Qi}.

Due to its simplicity and rich topological structure, the SSH model has been extensively studied and generalized \cite{Mar, Zha, Eli, Bid}. Extended SSH models with three or more sites per unit cell exhibit additional topological phases and edge states, depending on symmetry. For example, chiral edge states have been reported in inversion-symmetry-broken phases \cite{Mar}, and realizations of such models in optical and photonic systems have led to extensions into 2D structures with topological corner states \cite{Zha}. Models with periodic hopping modulations of period four, analyzed using Chebyshev polynomials, have revealed multiple distinct edge phases supported by numerical simulations \cite{Eli}.

A more recent class of 1D lattice models—termed mosaic lattices—was introduced in \cite{Wan}, inspiring numerous studies of their physical properties \cite{Zen1, Zen2, Gon, Dwi, Ngu1, Ngu2}. In these models, spatial modulation is introduced in either the on-site potential or the hopping amplitude, with periodic or quasiperiodic profiles. In particular, the off-diagonal quasiperiodic mosaic lattice with modulated hopping amplitudes was shown to host topologically nontrivial phases, supporting both zero- and nonzero-energy edge states when the modulation is commensurate with the lattice \cite{Zen2}. In contrast, incommensurate modulations induce Anderson localization \cite{And,Lee,Eve}, characterized by the absence of diffusion caused by destructive wave interference in a disordered potential, arising here from quasiperiodic off-diagonal disorder. In \cite{Zen1}, mosaic modulation of the on-site potentials was used to define a diagonal mosaic lattice model, where the interplay between modulation and superconducting pairing produced rich spectral features. Furthermore, studies of disordered diagonal mosaic lattices revealed that eigenstates at certain discrete energies exhibit critical power-law localization, whereas all other states are exponentially localized \cite{Ngu1, Ngu2}.

In this work, we investigate the edge states of an off-diagonal mosaic lattice model, where nearest-neighbor hopping amplitudes are modulated periodically at equally spaced intervals with period $\kappa$. We also consider the effects of disorder, allowing the hopping amplitudes $\beta$ to fluctuate randomly at intervals of $\kappa$. This model extends the off-diagonal quasiperiodic mosaic lattice of \cite{Zen2} by replacing quasiperiodic modulations with periodic or random ones, leading to several novel features. We show analytically that edge states emerge at discrete energy eigenvalues
\begin{equation}
E = \epsilon + 2t \cos\left( \frac{\pi}{\kappa} i \right), \quad i =1, \cdots, \kappa - 1, \label{eq:eigen}
\end{equation}
where $\epsilon$ and $t$ denote the on-site potential and (unmodulated) hopping amplitude, respectively. Numerical simulations reveal that these edge states exhibit robust localization and characteristic nodal structures, similar to those of critical states in disordered diagonal mosaic lattices \cite{Ngu1, Ngu2}, with their properties strongly dependent on the edge configurations of long and short bonds. Using the inverse participation ratio (IPR) as a measure of localization, we demonstrate that these topological edge states remain robust over a broad range of disorder strengths. In contrast, the additional $\beta$-dependent edge states that arise for
$\kappa \ge 4$
are fragile and vanish under relatively weak disorder.

Experimental realizations of 1D topological phases have been achieved in a variety of platforms, including ultracold atoms \cite{Ata, He}, photonic crystals \cite{Kra, Ver1, Ver2}, electronic systems \cite{Zhu, Nin}, and non-Hermitian structures \cite{Pan, Xia}.
More recently, photonic metamaterials have emerged as versatile testbeds for such models \cite{pmreview}. In particular, microwave split-ring resonator (SRR) chains have been used to implement the SSH model, demonstrating robust edge states even in the presence of disorder \cite{Jiang18}. This approach has been further extended to higher-order dimerized structures, such as trimerized and tetramerized chains, where topological boundary states have been systematically observed and characterized \cite{Liu21,Zhang21}. Beyond proof-of-concept demonstrations, these boundary states have also been harnessed for practical applications, including robust wireless energy transfer and wireless sensing \cite{Zhang21b,Song21}. These advances suggest that the off-diagonal mosaic lattice studied here could be readily emulated in SRR-based photonic platforms, enabling experimental verification of the theoretical predictions presented in this work.

\section{\label{sec:level1} Theoretical model}

We consider a standard tight-binding Hamiltonian for spinless electrons on a 1D lattice:
\begin{align}
H=\sum_{i}\epsilon_{i}c^{\dagger}_{i}c_{i}+\sum_{\langle i,j\rangle}\left(t_{ij}c^{\dagger}_{i}c_{j}+t_{ij}^*c^{\dagger}_{j}c_{i}\right),
\label{equation1}
\end{align}
where
$c^{\dagger}_{i}$ and $c_{i}$ are the creation and annihilation operators at site $i$,
$\epsilon_i$ is the on-site potential, and
$t_{ij}$ ($=t_{ji}^*$) is the hopping amplitude from site $j$ to site $i$.
The sum
$\langle i,j\rangle$ runs over distinct site pairs. Electron-electron interactions are neglected.

From Eq.~\eqref{equation1}, the discrete time-independent Schrödinger equation is
\begin{equation} \epsilon_{i} \psi_{i} + \sum_{j} t_{ij} \psi_{j} = E \psi_{i}, \label{equation2} \end{equation}
where
$\psi_i$ is the wave function amplitude at site $i$ and $E$ is the energy.
Assuming nearest-neighbor hopping and uniform on-site potential $\epsilon_i=\epsilon$, Eq.~\eqref{equation2} reduces to
\begin{equation} \epsilon \psi_{i} + t_{i-1} \psi_{i-1} + t_{i} \psi_{i+1} = E \psi_{i}, \label{equation3} \end{equation}
with $t_i\equiv t_{i,i+1}\in \mathbb{R}$ the hopping amplitude between sites $i$ and $i+1$.

In the off-diagonal mosaic lattice model \cite{Zen2}, hopping amplitudes are modulated as
\begin{equation}
t_{i} = \begin{cases} t\beta_{i}, & \text{if } i = j\kappa-m, \\ t, & \text{otherwise}, \end{cases} \label{equation4}
\end{equation}
where
$\kappa>1$ sets the mosaic modulation period, $j=1,2,\cdots,J$, and
$m\in [0,\kappa-1]$ determines the edge configuration. The modulation factor
$\beta_i$ may be a constant (periodic case) or a random variable (disordered case). For purely off-diagonal systems, we set $\epsilon=0$ in Eq.~(\ref{equation3}).

In experimental realizations, the parameters of our model have direct physical counterparts. The hopping amplitude
$t$ corresponds to the tunneling strength between adjacent potential wells in optical lattices or to the coupling between neighboring resonators in photonic waveguide arrays. The parameter
$\kappa$, which sets the spatial period of the mosaic modulation, can be implemented through superlattice potentials in ultracold atomic systems, by tailoring coupling strengths in photonic lattices, or by designing artificial electronic lattices. We also define
$\beta_i$ as the site-dependent bond parameter that specifies the relative strength of short and long bonds across the lattice. Experimentally, $\beta_i$ can be tuned by varying lattice depths in optical traps, adjusting resonator separations in photonic systems, or controlling orbital overlap in solid-state platforms. These correspondences illustrate that the theoretical formulation adopted here is directly relevant to experimentally accessible systems.

This study has two main objectives. First, we investigate the emergence of topological edge states in off-diagonal mosaic lattices, generalizing earlier results from extended SSH models \cite{Mar,Zha,Eli,Bid}. We show analytically that for arbitrary $\kappa$, edge states appear at energies given by Eq.~(\ref{eq:eigen}).
Second, we examine the robustness of these edge states under disorder, a topic largely unexplored in the context of extended SSH-type models. Our results demonstrate that off-diagonal mosaic lattices support topological edge states that remain robust even under strong, spatially varying hopping modulations.

\section{\label{sec:level1} Multi-band structure and edge states in finite lattices}

We analyze the periodic off-diagonal mosaic lattice with uniform modulation, $\beta_i = \beta$, focusing on the regime $\beta > 1$. For representative values of the inlay parameter $\kappa = 2$, 3, and 4, we demonstrate the emergence of edge states, with the results naturally extending to arbitrary integers $\kappa \ge 2$. We show that, for any such $\kappa$, edge states consistently emerge at all energy levels specified by Eq.~(\ref{eq:eigen}).

\subsection{\label{sec:level2} Two-band model ($\kappa=2$)}

\begin{figure}
\centering
\includegraphics[width=\columnwidth]{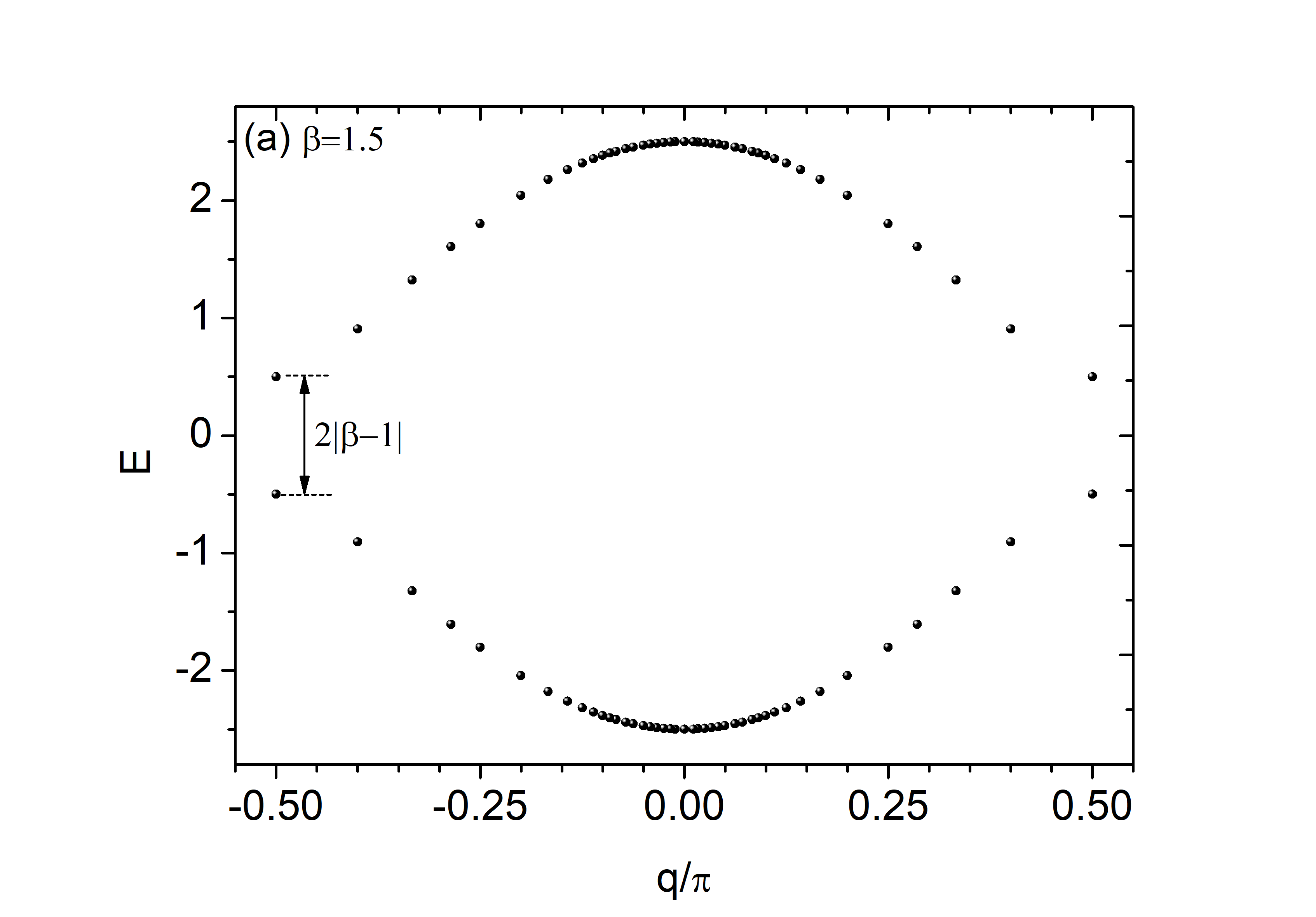}
\includegraphics[width=\columnwidth]{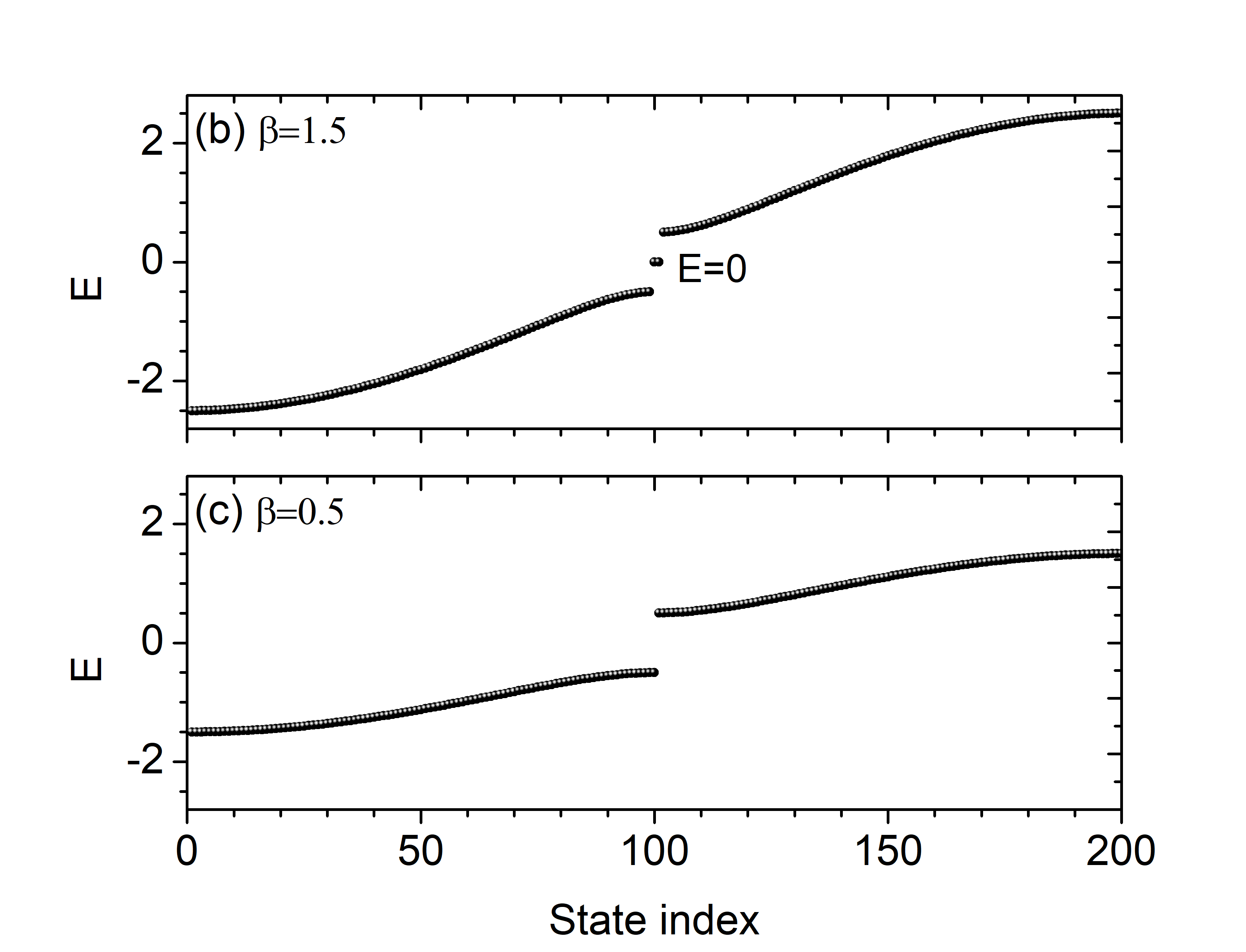}
\caption{(a) Energy $E$ versus wavenumber $q$ for $\kappa=2$ and $\beta = 1.5$ under periodic boundary conditions. A band gap opens at $q = \pm\pi/2$ when $\beta \ne 1$.
(b, c) Energy $E$ versus state index under open boundary conditions for $\beta = 1.5$ and $\beta = 0.5$, with $N = 200$ and $m = 0$. Two zero-energy edge states, localized at the left and right boundaries, appear for $\beta = 1.5$ but are absent for $\beta = 0.5$.}
\label{fig1}
\end{figure}

For $\kappa=2$, the periodic off-diagonal mosaic model reduces to the well-known 1D SSH model, which describes spinless fermions on a 1D lattice with alternating strong and weak hopping amplitudes \cite{SSH1}. This model is a paradigmatic example of a 1D topological insulator, illustrating bulk–boundary correspondence and the emergence of topological edge states.

According to Eq.~(\ref{equation4}), when $m = 0$, the first bond has hopping amplitude $t$. For an even number of sites, $\beta > 1$ corresponds to weak coupling between the edge and the bulk, resulting in a nontrivial topological phase with edge states, while $0 < \beta < 1$ leads to strong edge coupling and a trivial phase \cite{Stj}. A topological phase transition occurs at $\beta = 1$, where the band gap closes.

In the absence of on-site potential ($\epsilon = 0$), the Bloch Hamiltonian for $\kappa = 2$ under periodic boundary conditions is
\begin{align}
H_{\kappa=2} =
 t\begin{pmatrix}
  0 && 1+\beta e^{-2iq}  \\
  1+\beta e^{2iq} && 0
 \end{pmatrix},
\label{equation6}
\end{align}
yielding the dispersion relation
\begin{align}
\tilde{E}(q) = \pm \sqrt{1 + \beta^2 + 2\beta\cos(2q)},
\label{equation7}
\end{align}
where $\tilde{E} = E/t$ and $q \in [-\pi/2, \pi/2]$. A band gap of width $\Delta \tilde{E} = 2|\beta - 1|$ opens at $q = \pm \pi/2$, separating two topologically distinct phases. For $\beta > 1$, zero-energy edge states appear under open boundary conditions, whereas no edge states exist for $0 < \beta < 1$ \cite{Asb}. Hereafter, we set $t = 1$ and use $\tilde{E}$ and $E$ interchangeably when no confusion arises.

As shown in ~\cite{Eli}, when the number of sites $N$ is even, a pair of degenerate zero-energy edge states emerges in the nontrivial phase ($\beta > 1$), while all states remain extended in the trivial phase. At the critical point $\beta = 1$, the hopping amplitudes are uniform, the gap closes, and the system becomes metallic.

Figure~\ref{fig1} shows the numerically computed energy spectrum for $N = 200$ and $m = 0$, with $\beta = 1.5$ and $0.5$. In the topologically nontrivial case, two zero-energy edge states are clearly visible. For odd $N$, only a single edge state appears, localized at the left boundary.

\begin{figure}
\centering
\includegraphics[width=\columnwidth]{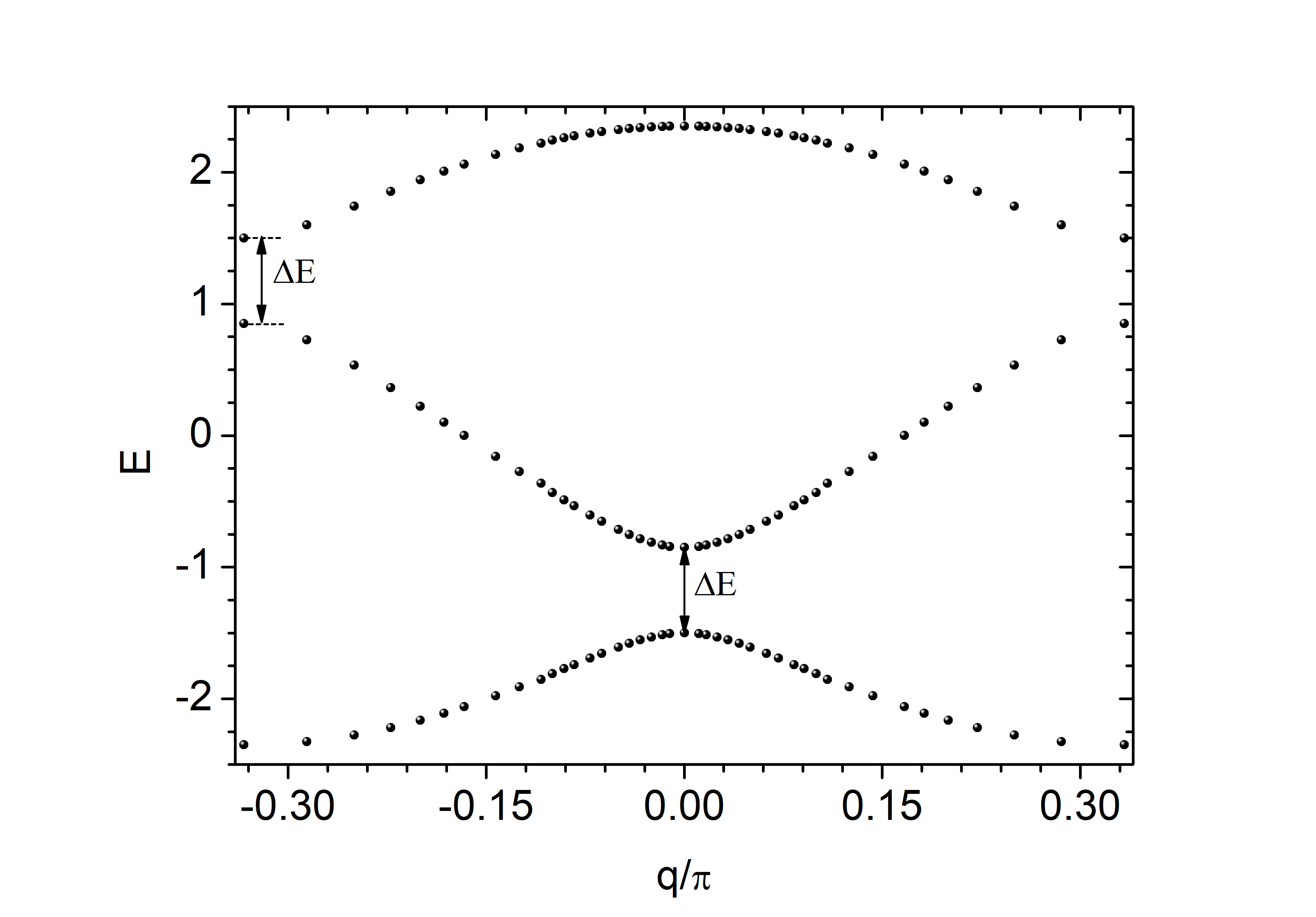}
\caption{
Energy $E$ versus wavenumber $q$ ($-\pi/3 \leq q \leq \pi/3$) for $\kappa=3$ and $\beta = 1.5$ under periodic boundary conditions.
For nonzero $\beta$, band gaps open at $q = 0$ and $q = \pm \pi/3$. }
\label{fig2}
\end{figure}

\begin{table*}
	\caption{\label{tab:table3} Edge states for $\kappa = 3$ with $\beta > 1$. Here, $m$ is defined in Eq.~(\ref{equation4}) and $J$ is an integer. For each configuration, the numbers of long bonds at the left and right edges are specified.}
		\centering\begin{tabular}{cccccccc}
        \hline
		&Case no.	&$m$ & No. of sites &Left long bonds &Right long bonds &Left edge states &Right edge states\\
			\hline
		&1	&0 & $3J$ &2 &2 &$E=\pm1$ &$E=\pm1$ \\
		&2	&0 & $3J+2$ &2 &1 &$E=\pm1$ & None \\
		&3	&0 & $3J+1$ &2 &0 &$E=\pm1$ & None \\
		&4	&1 & $3J+1$ &1 &1 &None &None \\
		&5	&1 & $3J$ &1 &0 &None &None \\
		&6	&2 & $3J+2$ &0 &0 &None &None \\
            \hline
		\end{tabular}
        \label{table1}
\end{table*}

\begin{figure}
\centering
\includegraphics[width=\columnwidth]{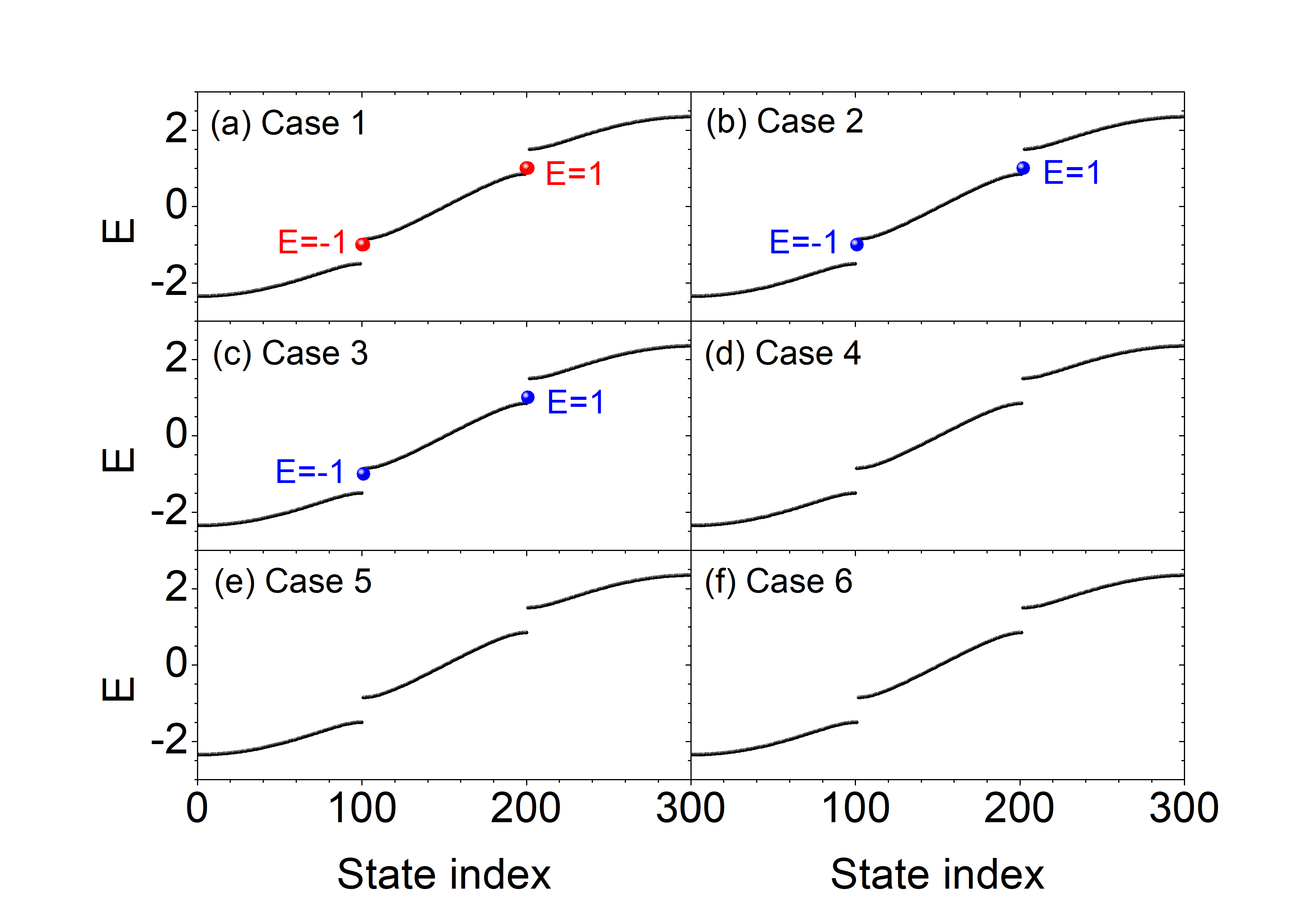}
\caption{Energy $E$ versus state index for $\kappa = 3$ and $\beta = 1.5$ under open boundary conditions. The six configurations listed in Table~\ref{table1} are shown separately. Red dots indicate doubly degenerate edge states localized at both boundaries, while blue dots denote single edge states localized at the left boundary.}
\label{ff3}
\end{figure}

\subsection{\label{sec:level2} Three-band model ($\kappa=3$)}

In this subsection, we analyze the case
$\kappa=3$. The spectrum of the model is determined by the eigenvalues of the Hamiltonian
\begin{align}
H_{\kappa=3} =
 t\begin{pmatrix}
 0 && 1 && \beta e^{-3iq} \\
 1 && 0 && 1 \\
 \beta e^{3iq} && 1 && 0 \\
 \end{pmatrix}.
\label{equation10}
\end{align}
Diagonalizing this matrix yields the dispersion relation (with $t=1$):
\begin{align}
E^3-\left(\beta^2+2\right)E-2\beta\cos(3q)=0,
\label{equation11}
\end{align}
from which the band structure reduces to
\begin{align}
E(q)=&\sqrt{\frac{4a}{3}}\cos\left[\frac{\arccos(b)}{3}\right],~\sqrt{\frac{4a}{3}}\cos\left[\frac{\arccos(b)}{3}\mp \frac{2\pi}{3}\right],
\label{equation12}
\end{align}
where
\begin{align}
a=\beta^2+2,~~
b=\left(\frac{3}{a}\right)^{3/2}\beta\cos(3q).
\label{equation13}
\end{align}

The band structure consists of three bands separated by gaps, forming asymmetric positive and negative branches. Band gaps open when $\beta \neq 1$. For example, for $\beta=1.5$, numerical results (Fig.~\ref{fig2}) show band gaps opening at $q=0$ and $\pm\pi/3$, each with width $\Delta E \simeq 0.65$.

In finite lattices, edge states appear at $E=\pm 1$ (for $\epsilon=0$), and their existence is determined by the specific arrangement of long (weak) and short (strong) bonds at the edges. Using the method in \cite{Eli}, we identify six independent edge configurations, summarized in Table~\ref{table1}. Edge states appear only when at least one edge contains two long bonds. Numerical spectra for cases (a)–(f) (with $N=300, 302, 301, 301, 300$, and 302 sites, respectively) are shown in Fig.~\ref{ff3}. Red dots indicate pairs of degenerate edge states localized at both edges, while blue dots mark single edge states on the left edge. These results align perfectly with Table~\ref{table1}.

The left-edge wave functions exhibit nodes at sites $i=3j$ ($j=1,2,\dots$), positioned immediately to the left of the short bonds ($t\beta$). The right-edge wave functions display mirror-symmetric nodes at sites immediately to the right of the short bonds.
These nodal structures are not accidental but intrinsic features of the topological edge states, as will be demonstrated in Sec.~\ref{sec:level22}.
Representative wave functions at $E=1$ for Case 1 are shown in Figs.~\ref{ff6}(a) and \ref{ff6}(b).

\begin{figure}
\centering
\includegraphics[width=\columnwidth]{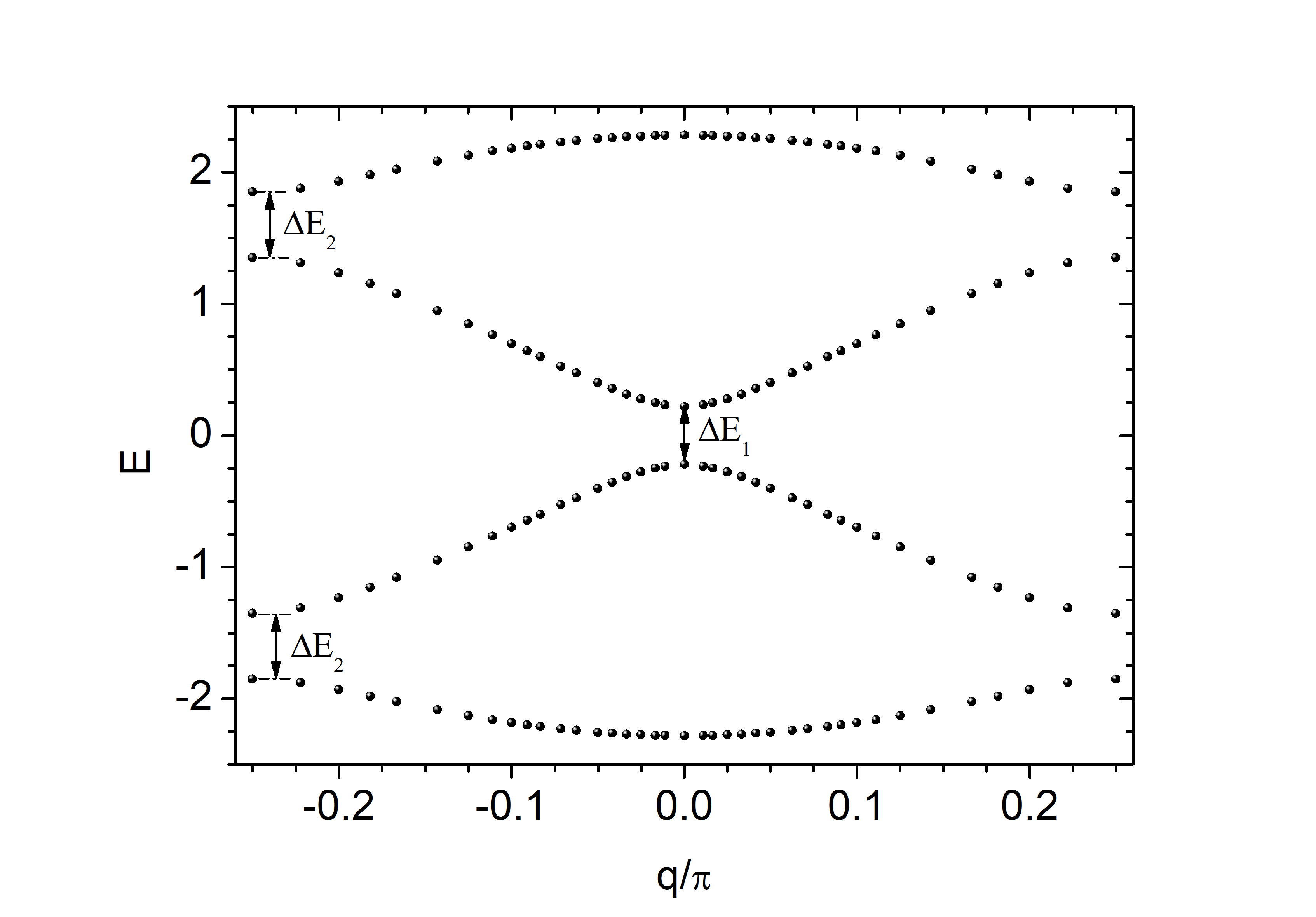}
\caption{Energy $E$ versus wavenumber $q$ ($-\pi/4 \leq q \leq \pi/4$) for $\kappa=4$ and $\beta = 1.5$ under periodic boundary conditions.
For nonzero $\beta$, band gaps open at $q = 0$ and $q = \pm \pi/4$.}
\label{ff4}
\end{figure}

\begin{table*}
	\caption{\label{tab:table3} Edge states for $\kappa = 4$ with $\beta > 1$. Here, $m$ is defined in Eq.~(\ref{equation4}) and $J$ is an integer. For each configuration, the numbers of long bonds at the left and right edges are specified.}
		\centering\begin{tabular}{cccccccc}
        \hline
		&Case no.	&$m$ & No. of sites &Left long bonds &Right long bonds &Left edge states &Right edge states\\
			\hline
		&1	&0 & $4J$ &3 &3 &$E=0, ~{\pm{\sqrt{2}}}$ &$E=0, ~{\pm\sqrt{2}}$ \\
		&2	&0 & $4J+3$ &3 &2 &$E=0, ~{\pm\sqrt{2}}$ & None \\
		&3	&0 & $4J+2$ &3 &1 &$E=0, ~{\pm\sqrt{2}}$ & $E=0$ \\
		&4	&0 & $4J+1$ &3 &0 &$E=0, ~{\pm\sqrt{2}}$ &$E={\pm\sqrt{1+\beta^2}}$ \\
		&5	&1 & $4J+2$ &2 &2 &None &None \\
        &6    &1 & $4J+1$ &2 &1 &None &$E=0$ \\
		&7	&1 & $4J$ &2 &0 &None &$E={\pm\sqrt{1+\beta^2}}$ \\
		&8	&2 & $4J$ &1 &1 &$E=0$ &$E=0$ \\
        &9    &2 & $4J+3$ &1 &0 &$E=0$ &$E={\pm\sqrt{1+\beta^2}}$\\
        &10    &3 & $4J+2$ &0 &0 &$E={\pm\sqrt{1+\beta^2}}$ &$E={\pm\sqrt{1+\beta^2}}$\\
        \hline
		\end{tabular}
        \label{table2}
\end{table*}

\begin{figure*}
\centering
\includegraphics[width=0.6\paperwidth]{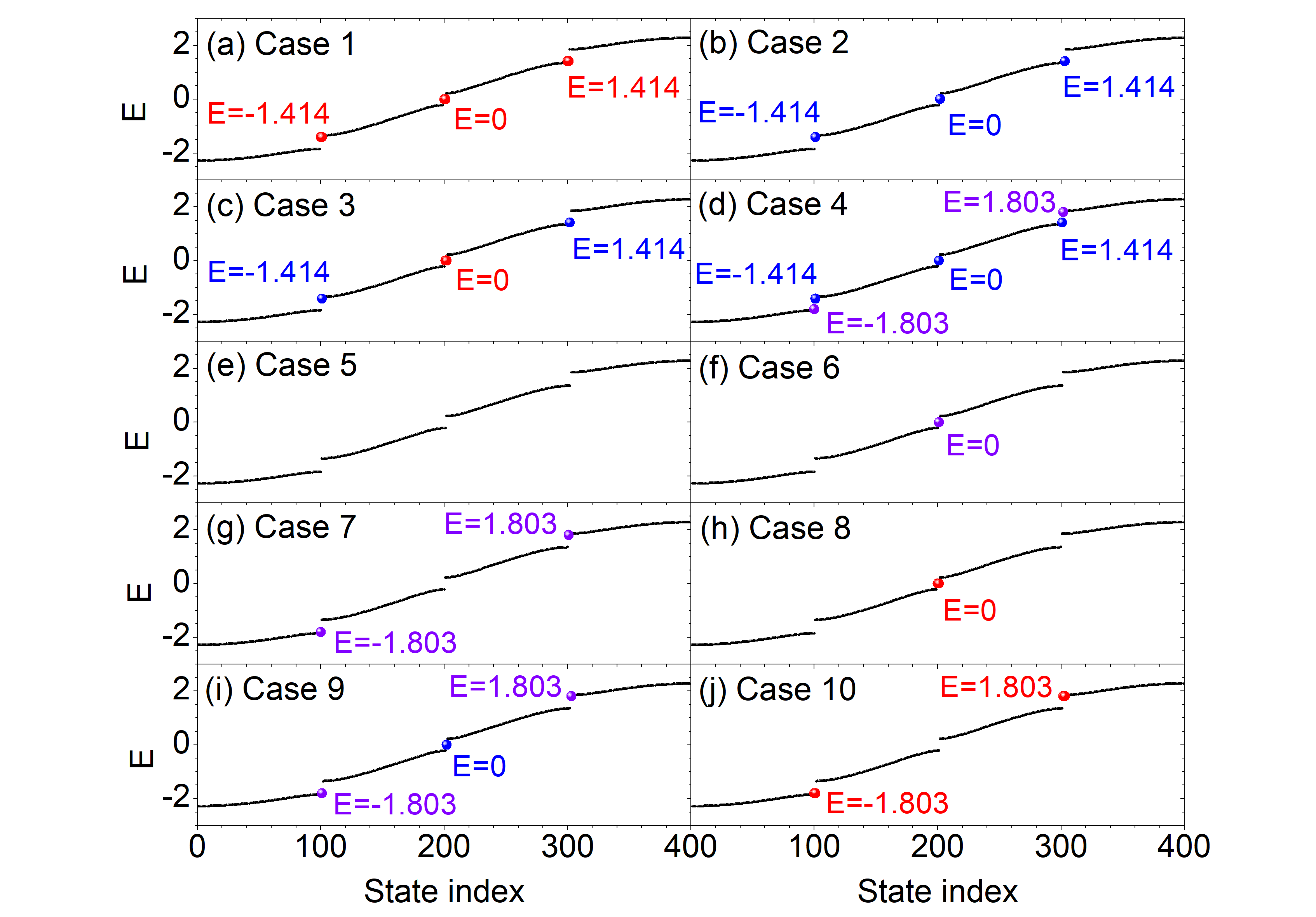}
\caption{Energy $E$ versus state index for $\kappa = 4$ and $\beta = 1.5$ under open boundary conditions. The ten configurations listed in Table~\ref{table2} are shown separately. Red dots indicate doubly degenerate edge states localized at both boundaries, while blue and violet dots represent single edge states localized at the left and right boundaries, respectively.}
\label{ff5}
\end{figure*}

\begin{figure}
\centering
\includegraphics[width=\columnwidth]{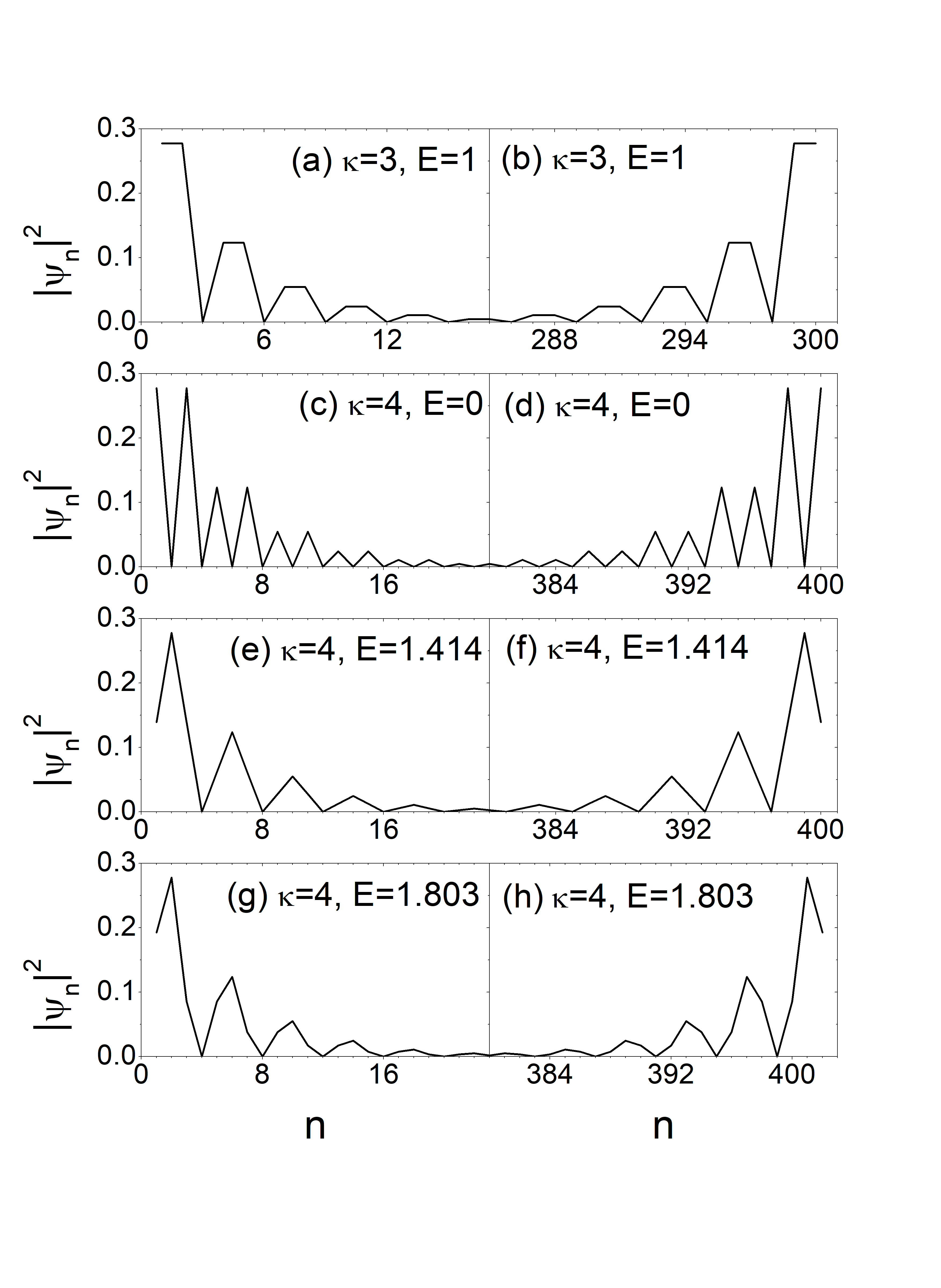}
\caption{Spatial distributions of eight selected edge-state wave functions for $\kappa = 3$ and $4$ with $\beta = 1.5$.
(a, b) States with $E = 1$ in Case 1 of Table~\ref{table1}, with total site number $N = 300$.
(c–f) States with $E = 0$ and $E = \sqrt{2}$ in Case 1 of Table~\ref{table2} for $N = 400$.
(g, h) States with $E = \sqrt{1 + \beta^2}$ in Case 10 of Table~\ref{table2} for $N = 402$.}
\label{ff6}
\end{figure}

\subsection{\label{sec:level21} Four-band model ($\kappa=4$)}

We next consider $\kappa=4$. The Hamiltonian is given by
\begin{align}
&H_{\kappa=4} =
 t\begin{pmatrix}
  0 && 1 && 0 && \beta e^{-4iq} \\
  1 && 0 && 1 && 0 \\
  0 && 1 && 0 && 1 \\
  \beta e^{4iq}  && 0 && 1 && 0
 \end{pmatrix}.
\label{equation101}
\end{align}
Diagonalizing this yields (with $t=1$):
\begin{align}
E^4-\left(\beta^2+3\right)E^2+1+\beta^2
-2\beta\cos(4q)=0,
\label{equation14}
\end{align}
and the band structure
\begin{align}
E(q)&=\sqrt{\frac{A\pm \sqrt{B}}{2}},~~-\sqrt{\frac{A\pm \sqrt{B}}{2}},
\label{equation15}
\end{align}
where
\begin{align}
    A=\beta^2+3,~~
    B=\beta^4+2\beta^2 +5+8\beta\cos(4q).
\end{align}
The spectrum has four symmetric bands separated by gaps. For $\beta=1.5$, band gaps open at $q=0$ and $q=\pm\pi/4$, with widths $\Delta E_1\simeq 0.44$ and $\Delta E_2\simeq 0.5$ (Fig.~\ref{ff4}).

Edge states in a finite lattice appear at $E=0$, $\pm \sqrt{2}$, and $\pm\sqrt{1+\beta^2}$ (for $\epsilon=0$). Their existence depends on the left and right edge configurations, which can be determined analytically using the results in \cite{Eli}. There are ten independent configurations defined by the number of long bonds at each edge, as summarized in Table~\ref{table2}. Edge states are present in all cases except for Case 5. Numerical energy spectra are shown in Fig.~\ref{ff5}, where red dots indicate two degenerate edge states localized at opposite edges, and blue and violet dots denote nondegenerate edge states at the left and right edges, respectively. The total number of sites used for cases (a)–(j) are
$N=400$, 403, 402, 401, 402, 401, 400, 400, 403, and 402, respectively. These numerical results are in perfect agreement with Table~\ref{table2}.

Numerical results for representative edge-state wave functions are shown in Figs.~\ref{ff6}(c–h). For the eigenstates at
$E=\pm \sqrt{2}$, the left-edge wave functions exhibit nodes at all sites immediately to the left of the short bonds with hopping amplitude $t\beta$, while the right-edge wave functions display mirror-symmetric nodes at the corresponding sites immediately to the right of the short bonds. For the $E=0$ eigenstates, the left-edge wave functions have nodes both immediately to the left of the short bonds and at the midway sites between them, with the right-edge states again showing mirror-symmetric nodal structures at the corresponding right-edge positions. These nodal structures are intrinsic features of the topological edge states.

The edge states at $E=\pm\sqrt{1+\beta^2}$ require separate consideration. These states exist only for
$\kappa\ge 4$
and occur exclusively at the edge terminated with a short bond, as in Cases 4, 7, 9, and 10 of Table \ref{table2}. Their nodal patterns are distinctive: left-edge states have nodes shifted one site further left than those adjacent to the short bonds, while right-edge states exhibit mirror-symmetric nodes one site further right. These are termination-induced states, analogous to Tamm states \cite{tamm}, and are independent of the bulk spectrum. Since they are not connected to bulk topological invariants, there is no bulk–boundary correspondence and no symmetry protecting their energies. Consequently, they are fragile to disorder and disappear even under weak fluctuations of
$\beta$. The eigenvalues $E=\pm\sqrt{1+\beta^2}$ and their nodal properties
are derived in Appendix A.

Overall, these results confirm that topological edge states occur at the energy eigenvalues specified by Eq.~(\ref{eq:eigen}) for all $\kappa\ge 2$.
For $\kappa\ge 4$,
additional $\beta$-dependent edge states appear, but only when the edge terminates with a short bond, as summarized in Table \ref{table2}. The number and character of these states are highly sensitive to the detailed arrangement of long and short bonds at the lattice edges.

\subsection{\label{sec:level22} Topological edge states for general $\kappa$}

In this subsection, we prove that for any integer $\kappa\ge 2$, a finite off-diagonal mosaic lattice with open boundary conditions supports edge states at the energy eigenvalues given by Eq.~(\ref{eq:eigen}).
Our approach generalizes the formalism developed in \cite{Eli} for $\kappa=4$, which in turn builds on the theorem established in \cite{BGL}.

The model described by Eq.~(\ref{equation3}) is a special case of a general three-term recurrence relation with periodic coefficients, $t_{i+\kappa}=t_i$.
Following \cite{Eli}, we introduce
$p_i=(t_1t_2\cdots t_i)\psi_{i+1}$.
From Eq.~(\ref{equation3}), one readily derives
\begin{align}
p_i=\varepsilon p_{i-1}-t_{i-1}^2p_{i-2},
\label{eq:bgl}
\end{align}
where $\varepsilon\equiv E-\epsilon$.
Imposing the boundary condition $\psi_0=0$, which implies $p_{-1}=0$, brings Eq.~(\ref{eq:bgl}) into the form analyzed in \cite{BGL}. There it was shown that the solutions satisfy
\begin{align}
p_{i\kappa+s}=h^{i-1}p_{\kappa+s}U_{i-1}\left(\frac{w}{2h}\right)
-h^ip_sU_{i-2}\left(\frac{w}{2h}\right),
\label{eq:bgl2}
\end{align}
where $U_i$ denotes the Chebyshev polynomials of the second kind, and $w\equiv w(\varepsilon)$ is a degree-$\kappa$ polynomial
defined as
\begin{align}
w(\varepsilon)=\frac{p_{2\kappa-1}(\varepsilon)}{p_{\kappa-1}(\varepsilon)}.
\label{eq:bgl22}
\end{align}
Assuming all $t_i$ are positive real numbers, we define
\begin{align}
h=t_1t_2\cdots t_\kappa.
\end{align}
Substituting the definitions of $p_i$ and $h$ into Eq.~(\ref{eq:bgl2}) yields
\begin{align}
    \psi_{i\kappa+s}=\psi_{\kappa+s}U_{i-1}\left(\xi\right)-\psi_{s}U_{i-2}\left(\xi\right),
\label{eq:bgl3}
\end{align}
where $\xi\equiv w/(2h)$.

When the number of long bonds at the left edge reaches its maximum value, $\kappa-1$, edge states appear at all eigenvalues $E=\epsilon+2\cos(\pi i/\kappa)$, regardless of the bond configuration at the right edge, and no $\beta$-dependent edge states occur on the left edge.
This situation corresponds to Cases 1–3 in Table~1 for $\kappa=3$ and to Cases 1–4 in Table~2 for $\kappa=4$.
For the analytic proof of edge-state existence in these cases, it is most convenient to consider the configuration where the right edge contains $\kappa-2$ long bonds, since the condition in Eq.~(21) then takes a particularly simple form.
This corresponds to Case 2 in Tables~1 and 2 for $\kappa=3$ and $\kappa=4$, respectively.
In this configuration, the total number of lattice sites is $N=\kappa J +\kappa-1$, where $J$ is an arbitrary large integer.

For edge states to exist at the left boundary, the wave function at site $N+1=\kappa(J+1)$ must vanish. From Eq.~(\ref{eq:bgl3}), this condition becomes
\begin{align}
    \psi_{\kappa(J+1)}&=\psi_{2\kappa}U_{J-1}\left(\xi\right)-\psi_{\kappa}U_{J-2}\left(\xi\right)\nonumber\\
&=\psi_{\kappa}\left[\frac{\psi_{2\kappa}}{\psi_{\kappa}}U_{J-1}\left(\xi\right)-U_{J-2}\left(\xi\right)\right]=0.
\label{eq:bgl4}
\end{align}
In the present configuration of the periodic off-diagonal mosaic lattice, the hopping amplitudes are fixed as $t_1 = t_2 = \cdots = t_{\kappa-1} = 1$ and $t_\kappa = \beta$.
Using the periodicity of $t_i$ and Eq.~(\ref{eq:bgl22}),
we obtain
\begin{align}
\frac{\psi_{2\kappa}}{\psi_{\kappa}}=\frac{w}{t_1t_2\cdots t_\kappa}=2\xi.
\end{align}
Substituting this relation into Eq.~(\ref{eq:bgl4}) and applying the Chevyshev recursion relation,
\begin {align}
U_J(\xi)=2\xi U_{J-1}(\xi)-U_{J-2}(\xi),
\end{align}
gives $\psi_\kappa U_J(\xi)=0$.
Since $U_J(\xi)$ does not vanish identically, we conclude that $\psi_\kappa=0$, which
implies that the wave function exhibits nodes at all sites $i=\kappa j$ for $j=1, 2, \cdots$.

From Eq.~(\ref{equation3}), we also obtain the relations
\begin{align}
-\varepsilon \psi_1+\psi_2&=0,\nonumber\\
\psi_1-\varepsilon \psi_2+\psi_3&=0,\nonumber\\
& \vdots\nonumber\\
\psi_{\kappa-2}-\varepsilon \psi_{\kappa-1}+\psi_\kappa&=0,
\label{eq:crr}
\end{align}
which can be compactly written as
\begin{align}
\psi_{i-1}-\varepsilon\psi_i+\psi_{i+1}=0,\quad i=1,2,\cdots,\kappa-1.
\end{align}
By comparison with the recursion relation of Chebyshev polynomials $U_i(\varepsilon/2)$, we identify
$\psi_i=U_{i-1}(\varepsilon/2)$, which in particular gives $\psi_\kappa=U_{\kappa-1}(\varepsilon/2)$.
Since the zeros of $U_{n}(x)$ are located at $x=\cos[k\pi/(n+1)]$ for $k=1, 2, \cdots, n$,
we find that $\psi_\kappa=0$ when
\begin{align}
\frac{\varepsilon}{2}=\cos\left(\frac{k\pi}{\kappa}\right),\quad k=1,2,\cdots,\kappa-1.
\end{align}
Therefore, edge states exist when the energy satisfies $E=\epsilon+2\cos(\pi i/\kappa)$ with $i=1,2,\cdots,\kappa-1$.
Even when the number of long bonds at the right edge differs from $\kappa-2$, the left-edge wave functions continue to exhibit nodes at all sites $i=\kappa j$, lying immediately to the left of the short bonds. This property explains why the eigenvalues remain independent of $\beta$, ensuring robustness of the edge states against random variations of $\beta$.

Extending the recursion relations in Eq.~(\ref{eq:crr}) further, we obtain
\begin{align}
\psi_{\kappa-1}-\varepsilon \psi_\kappa+\beta\psi_{\kappa+1}=0,
\label{eq:crr2}
\end{align}
and
\begin{align}
\beta\psi_{\kappa}-\varepsilon \psi_{\kappa+1}+\psi_{\kappa+2}&=0,\nonumber\\
\psi_{\kappa+1}-\varepsilon \psi_{\kappa+2}+\psi_{\kappa+3}&=0,\nonumber\\
& \vdots\nonumber\\
\psi_{2\kappa-2}-\varepsilon \psi_{2\kappa-1}+\psi_{2\kappa}&=0.
\label{eq:crr3}
\end{align}
Using $\psi_\kappa=0$, Eq.~(\ref{eq:crr2}) decouples from Eq.~(\ref{eq:crr3}). The eigenvalues obtained from Eq.~(\ref{eq:crr3}) coincide with those from Eq.~(\ref{eq:crr}). Equation~(\ref{eq:crr2}) further shows that the edge-state wave function decays away from the boundary as a power of $-1/\beta$.
Thus, for arbitrary
$\kappa$, finite off-diagonal mosaic lattices host topological edge states whose eigenvalues are independent of
$\beta$ and whose nodal structures are determined solely by the underlying mosaic pattern, highlighting their inherent topological robustness.

\begin{figure}
\centering
\includegraphics[width=\columnwidth]{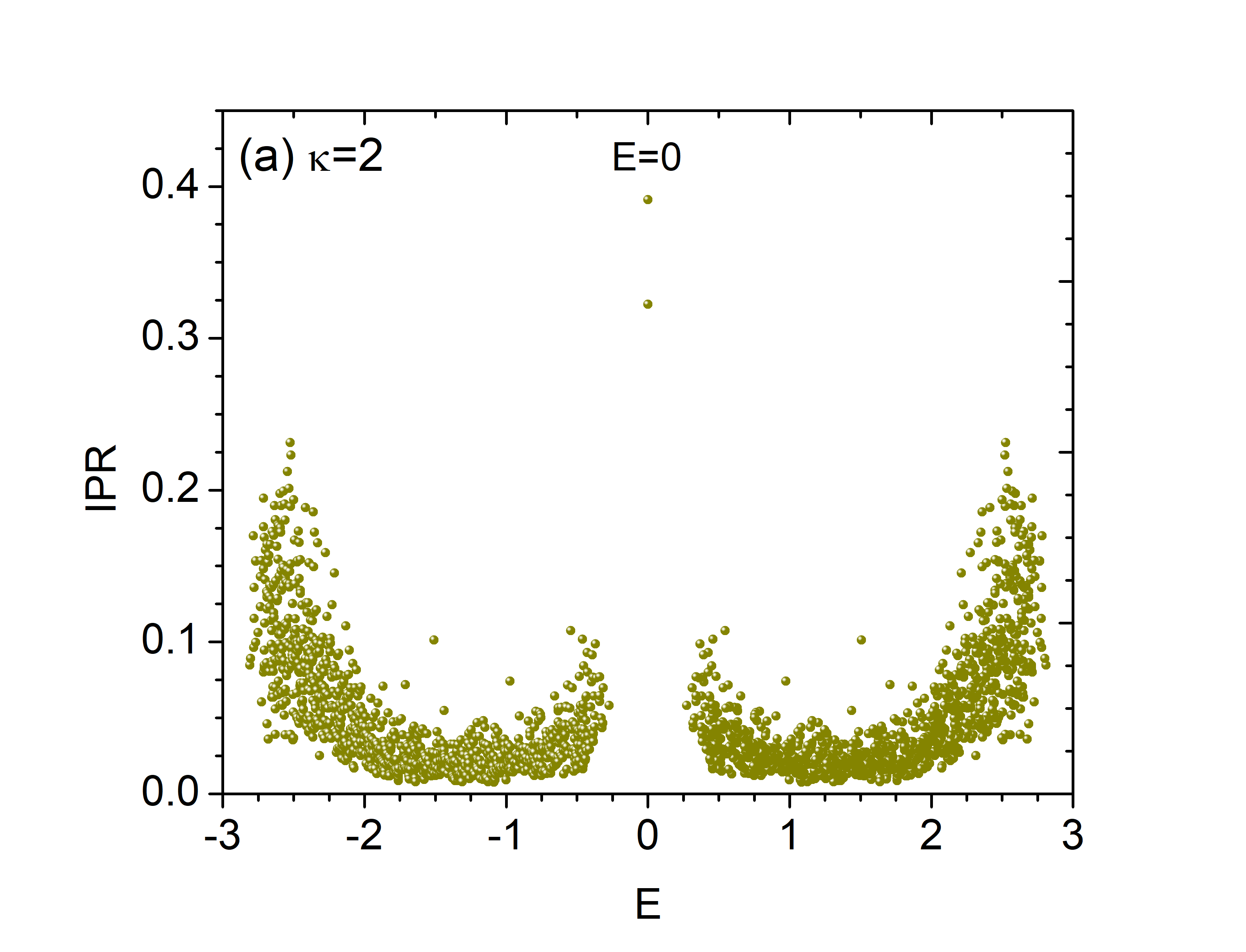}
\includegraphics[width=\columnwidth]{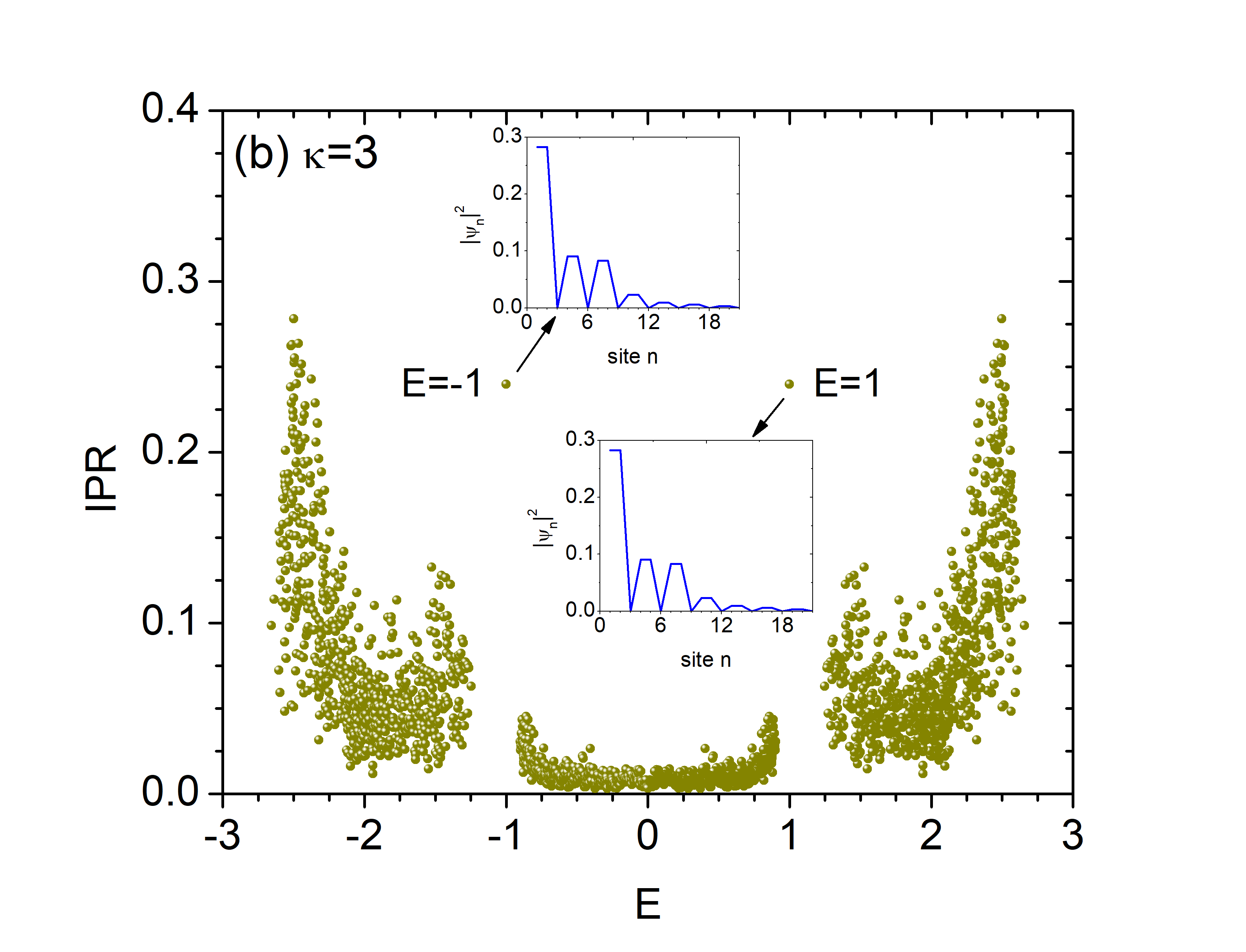}
\includegraphics[width=\columnwidth]{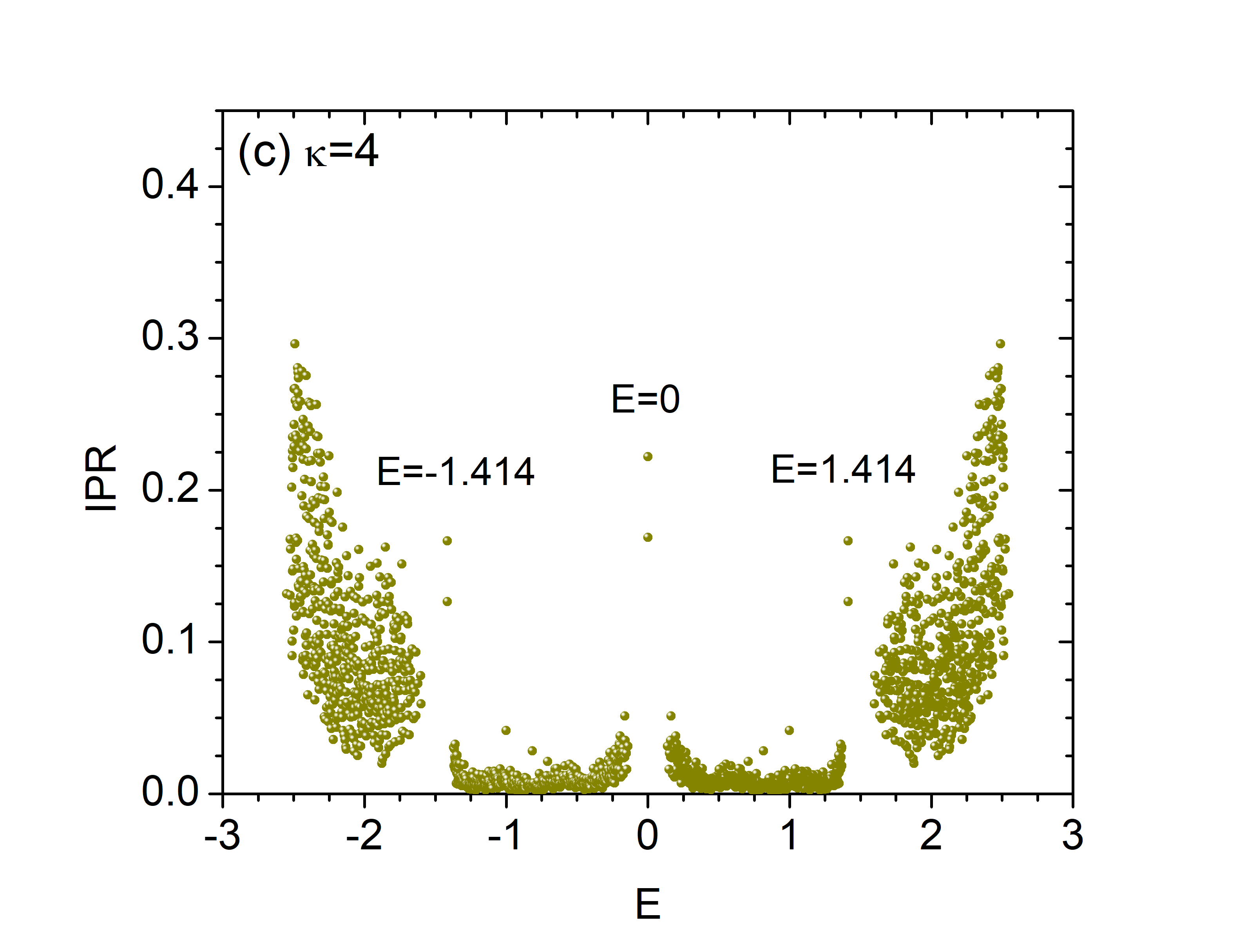}
\caption{Inverse participation ratio (IPR) as a function of energy $E$ for a single random realization of a chain of length $L = 2000$ under open boundary conditions, with $\beta_0 = 1.5$, $m=0$, and disorder strength $W = 1$. Results are shown for (a) $\kappa = 2$, (b) $\kappa = 3$ (Case 2 in Table~\ref{table1}), and (c) $\kappa = 4$ (Case 1 in Table~\ref{table2}). Edge states appear in pairs in (a) and (c), and individually in (b). The insets in (b) show that the node structure of the edge state remains robust against disorder.}
\label{ff7}
\end{figure}

\begin{figure}
\centering
\includegraphics[width=\columnwidth]{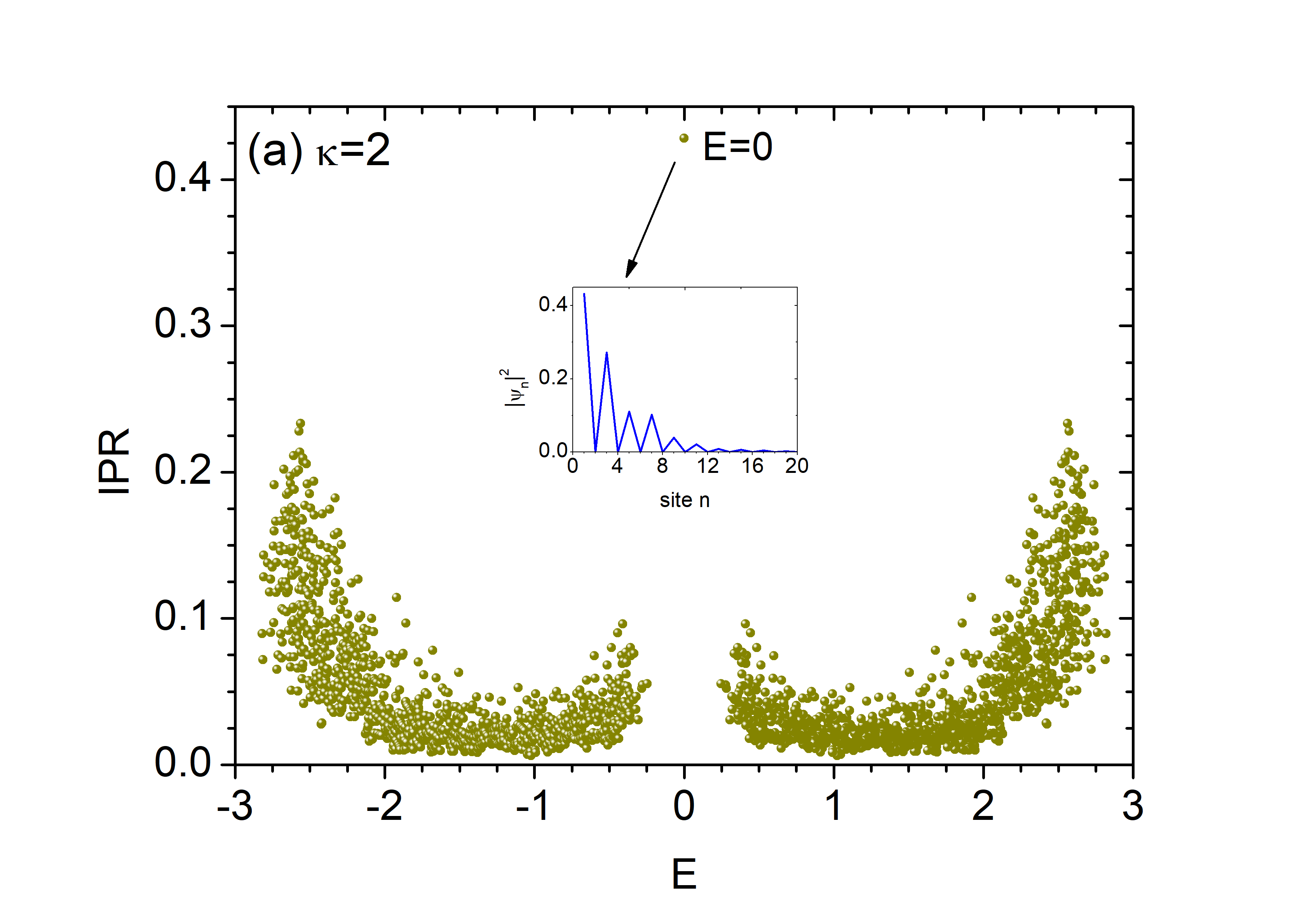}
\includegraphics[width=\columnwidth]{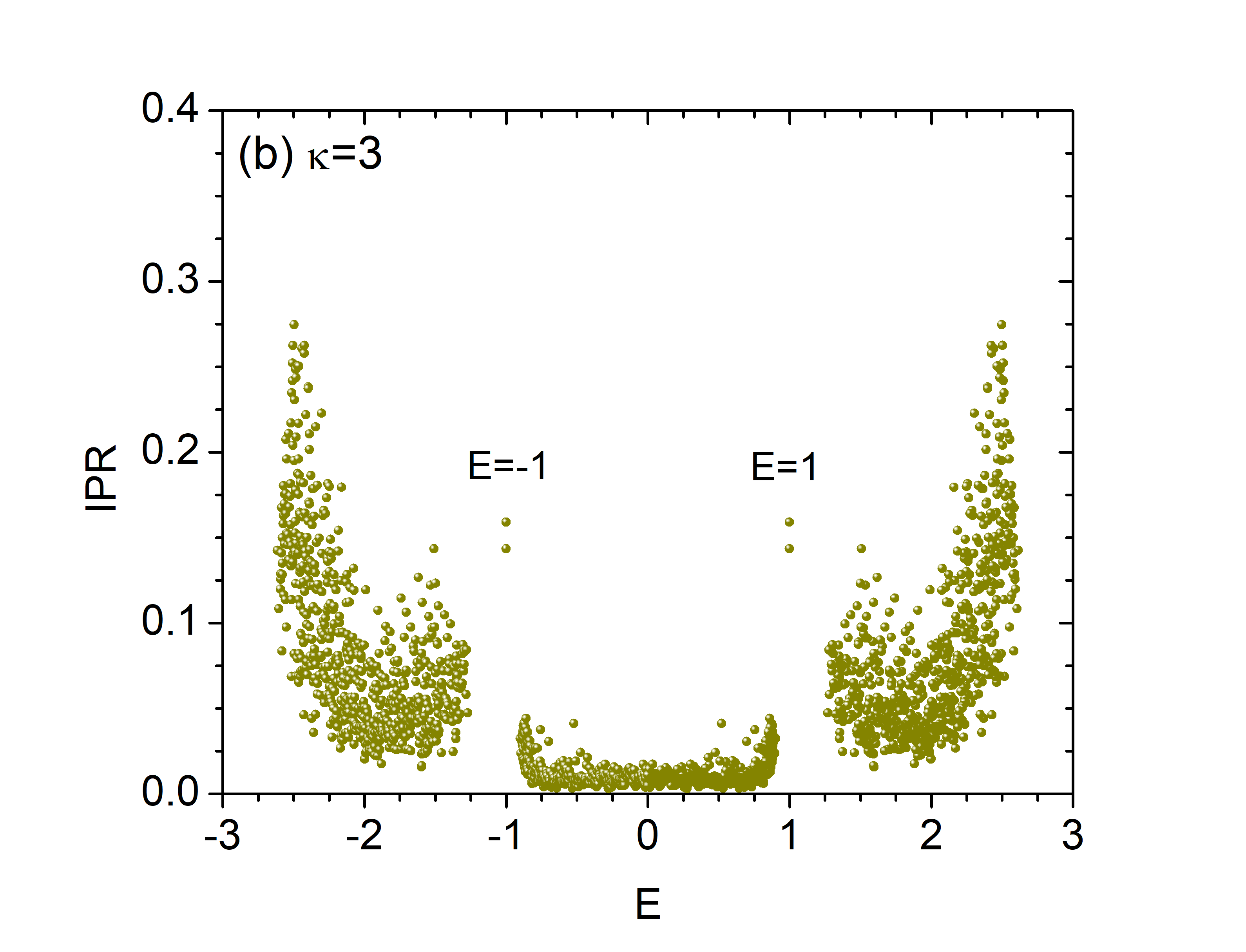}
\includegraphics[width=\columnwidth]{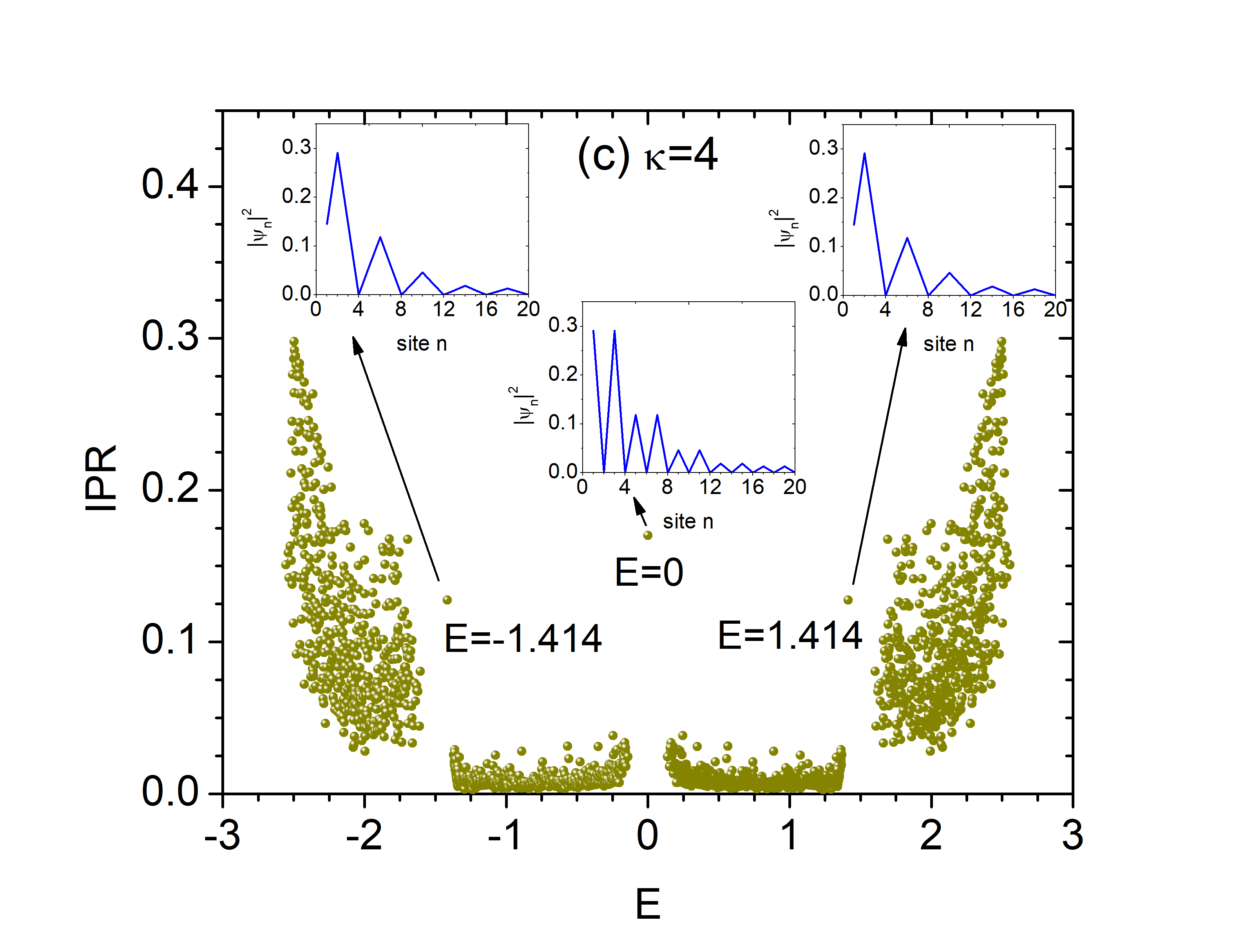}
\caption{IPR as a function of energy $E$ for a single random realization of a chain of length $L = 2001$ under open boundary conditions, with $\beta_0 = 1.5$, $m=0$, and disorder strength $W = 1$. Results are shown for (a) $\kappa = 2$, (b) $\kappa = 3$ (Case 1 in Table~\ref{table1}), and (c) $\kappa = 4$ (Case 4 in Table~\ref{table2}). Edge states appear in pairs in (b), and individually in (a) and (c). The insets in (a) and (c) show that the node structure of the topological edge state remains robust against disorder. In (c), however, edge states at $E = \pm \sqrt{1 + \beta^2}$ are absent due to the relatively strong disorder.}
\label{ff8}
\end{figure}

\begin{figure}
\centering
\includegraphics[width=\columnwidth]{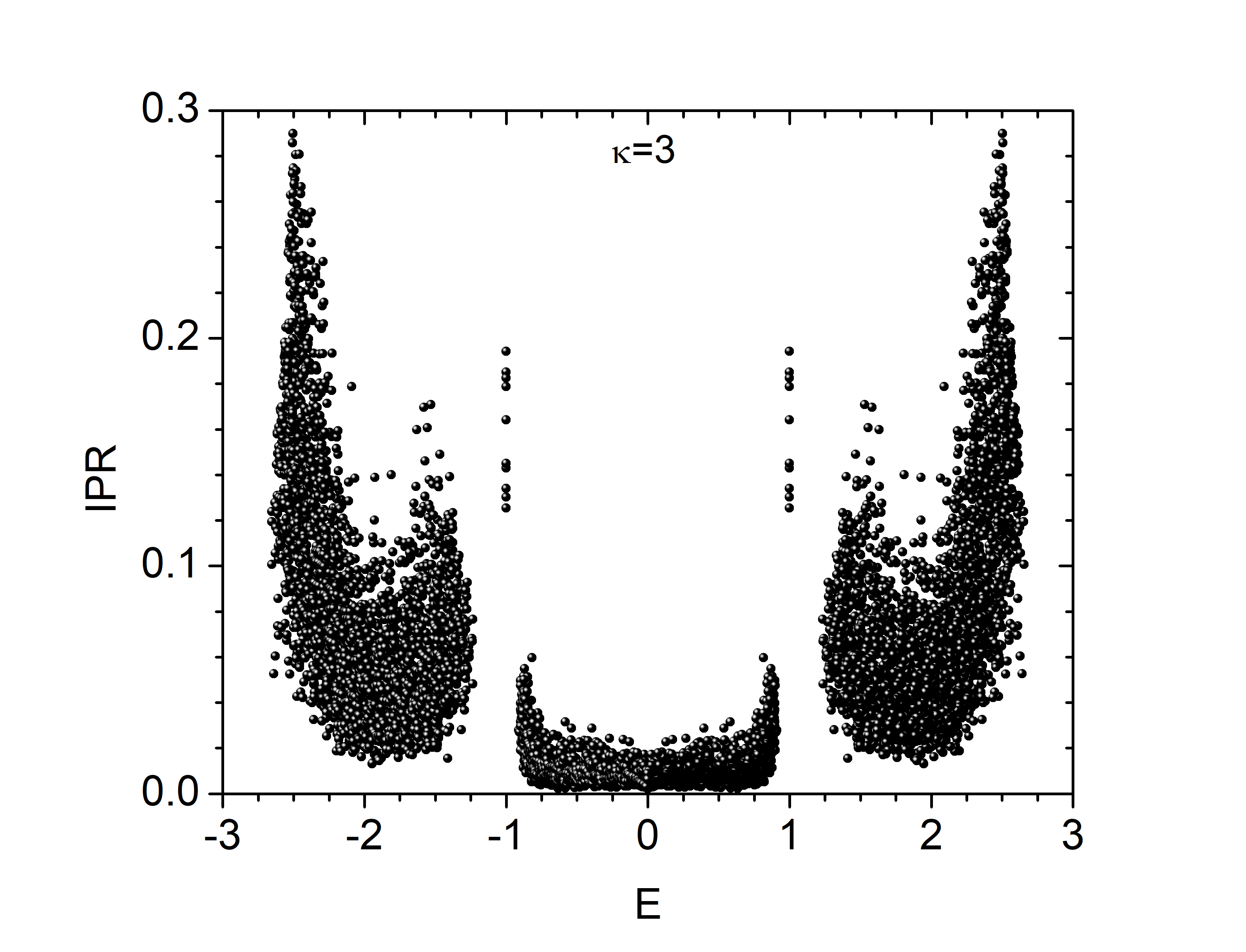}
\caption{IPR versus energy $E$ for a chain of length $L = 2001$ under open boundary conditions, with $\kappa=3$, $\beta_0 = 1.5$, $m=0$, and disorder strength $W = 1$. Data from five independent random realizations are overlaid. In all cases, edge states consistently appear in pairs at $E = \pm 1$.}
\label{ff9}
\end{figure}

\begin{figure}
\centering
\includegraphics[width=\columnwidth]{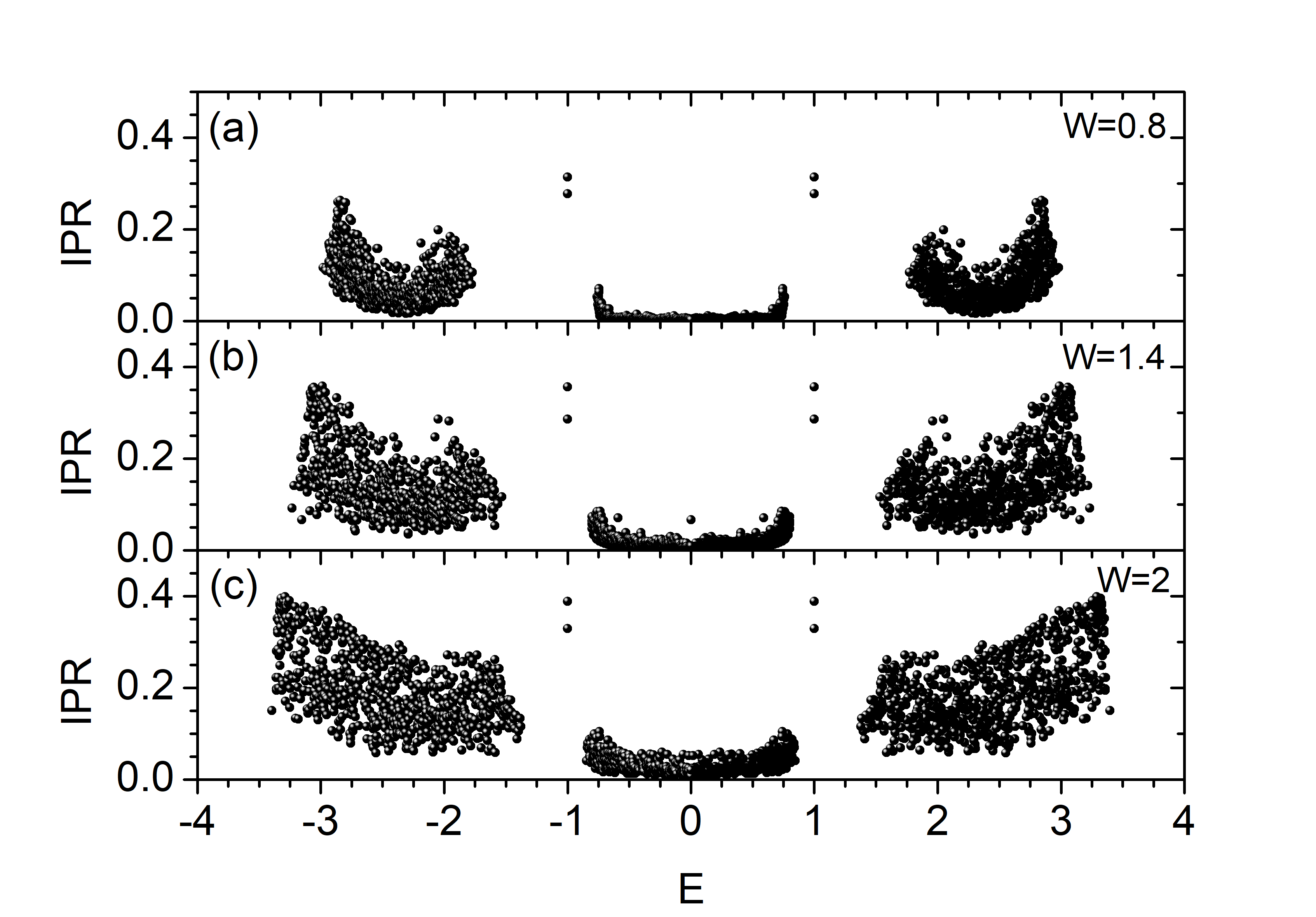}
\caption{IPR versus energy $E$ for a single random realization of a chain of length $L = 2001$ under open boundary conditions, with $\kappa = 3$, $\beta_0 = 2$, $m=0$, and disorder strengths $W = 0.8$, $1.4$, and $2$. In all cases, edge states appear in pairs at $E = \pm 1$.}
\label{ff10}
\end{figure}

\begin{figure}
\centering
\includegraphics[width=\columnwidth]{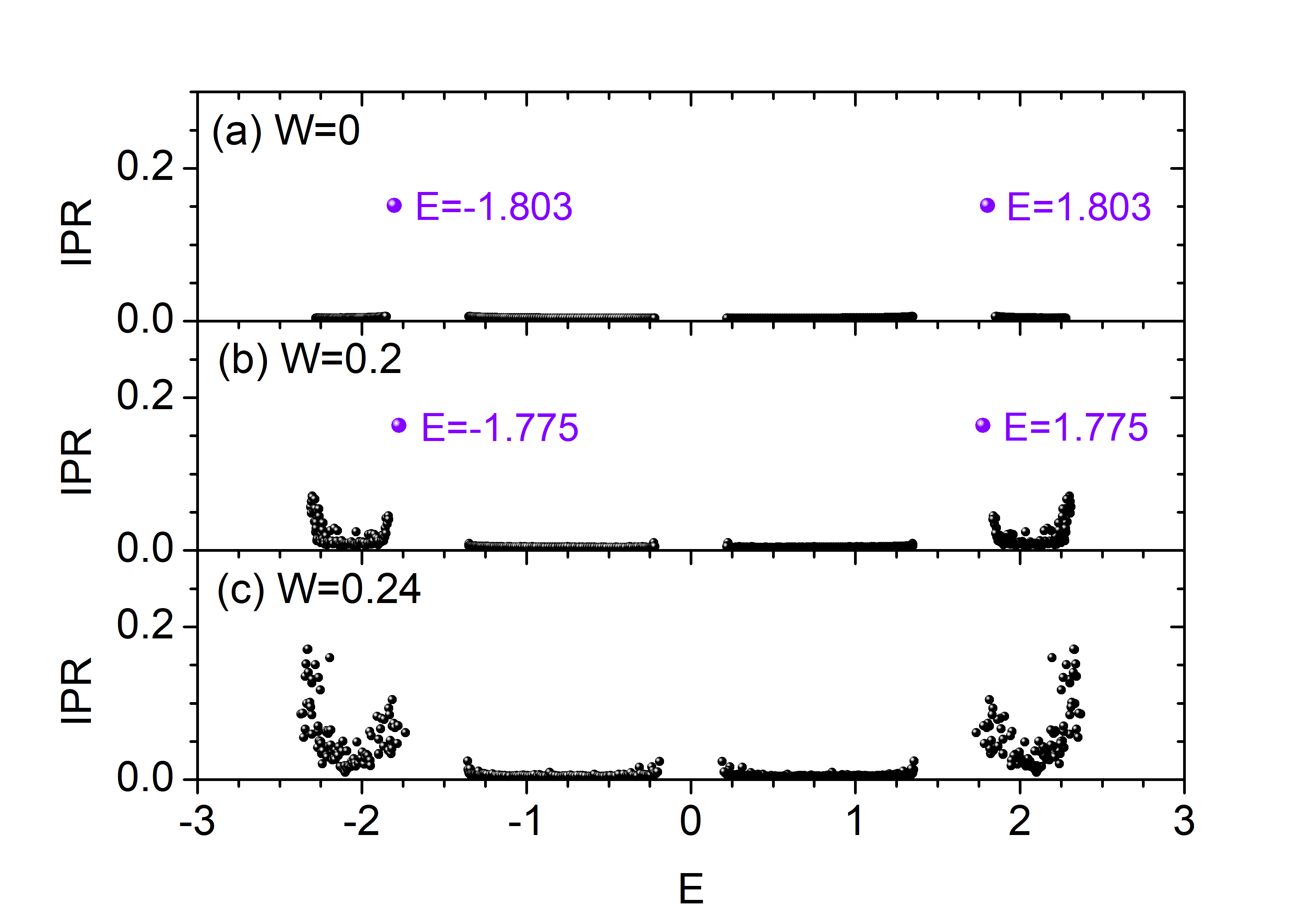}
\caption{IPR versus energy $E$ for a single random realization of a chain of length $L = 400$ under open boundary conditions, with $\kappa = 4$, $\beta_0 = 1.5$, $m = 1$, and disorder strengths $W = 0$, $0.2$, and $0.24$ (Case 7 in Table~\ref{table2}). In panels (a) and (b), edge states appear in pairs at $E \approx \pm\sqrt{1 + \beta^2}$, but they are absent in (c).}
\label{ff11}
\end{figure}

\section{\label{sec:level1} Robustness of edge states under disorder}

In the periodic off-diagonal mosaic model, our analysis shows that nearly all states correspond to extended bulk states, except for specific in-gap edge states that are exponentially localized at the boundaries of a finite chain. While numerous studies have demonstrated that zero-energy edge states are topologically protected against disorder \cite{Asb}, the stability of nonzero-energy edge states under disorder remains less well explored. To address this, we investigate their behavior within a disordered off-diagonal mosaic lattice chain, where
$\beta_i$ in Eq.~(\ref{equation4}) are independent random variables uniformly distributed in the interval $[\beta_0-W/2,\beta_0+W/2]$, with $\beta_0$ denoting the average value of $\beta_i$. The parameter $W$ characterizes the disorder strength, while all other hopping amplitudes are fixed at
$t=1$.

Before presenting the numerical analysis, we briefly discuss the $\kappa=1$ case without mosaic modulation. In this limit, it is well known that purely off-diagonal disorder leads to localization in a manner similar to diagonal disorder \cite{And, Lee, Eve}. However, near the band center ($E\rightarrow 0$), whether states remain localized or become extended remains a subject of debate \cite{Cher, Izra}. Except in this critical region, all states are exponentially localized for uncorrelated hopping amplitudes, regardless of the disorder strength. In contrast, introducing mosaic modulation ($\kappa\ge 2$) gives rise to distinct topological and localization features absent in conventional off-diagonal disordered systems, as we will demonstrate.

The degree of localization is quantitatively characterized by the IPR \cite{Weg}. For the $k$-th eigenstate, represented by  $\left(\psi_{1}^{(k)},\psi_{2}^{(k)},\cdots,\psi_{N}^{(k)}\right)^{\rm T}$ with eigenvalue $E_{k}$, the IPR is defined as
\begin{align}
{\rm IPR}\left(E_{k}\right)=\frac{\sum_{n=1}^{N}\left|\psi_{n}^{(k)}\right|^4}{\left(\sum_{n=1}^{N}\left|\psi_{n}^{(k)}\right|^2\right)^2}.
\label{equation18}
\end{align}
The IPR ranges from zero for completely extended states to unity for states fully localized in the limit of infinite system size. In finite systems, extended states typically yield ${\rm IPR}\sim O(1/L)$, while localized states exhibit ${\rm IPR}\sim O(1)$.

To compute the eigenvalues $E$ and eigenstates $\psi=\left(\psi_{1},\psi_{2},\cdots,\psi_{N}\right)^{\rm T}$ in the presence of off-diagonal disorder, we numerically solve the eigenvalue problem $\hat{H}\psi=E\psi$,
where $\hat{H}$ is the disordered off-diagonal mosaic Hamiltonian:
\begin{align}
\hat{H} =
 \begin{pmatrix}
  \epsilon & t_{1} & 0 & \cdots & 0 & 0 & 0\\
 t_{1} &  \epsilon & t_{2} & \cdots & 0 & 0 & 0\\
 0 & t_{2} &  \epsilon & \cdots & 0 & 0 & 0\\
 \vdots  & \vdots  & \vdots &  \ddots & \vdots & \vdots & \vdots\\
 0 & 0 & 0 & \cdots & \epsilon & t_{N-2} & 0\\
 0 & 0 & 0 & \cdots &  t_{N-2} & \epsilon & t_{N-1}\\
 0 & 0 & 0 & \cdots & 0 & t_{N-1} & \epsilon
 \end{pmatrix}.
\label{equation20}
\end{align}

Figures~\ref{ff7} and \ref{ff8} show the IPR as a function of energy $E$ for single random realizations in lattice chains of length $L = 2000$ and $L=2001$, with $\kappa=2$, 3, and 4 at $\beta_0=1.5$ and $W=1$.
In all cases, the edge configuration has $m=0$ and the first
$(\kappa-1)$ bonds on the left edge are long bonds. The examples for $\kappa=3$ and 4 correspond to selected configurations from Tables~\ref{table1} and \ref{table2}.
The numerical results reveal that all eigenstates are localized, including those far from the band edges, where the IPR values remain comparatively small. Two types of localized states are observed: edge states and bulk states.

The localization properties exhibit several notable features. First, the IPR spectra are arranged into distinct clusters separated by band gaps, in contrast to conventional off-diagonal disordered systems ($\kappa=1$), where such gaps are absent. As discussed earlier, the emergence of band gaps is a characteristic feature of the periodic off-diagonal mosaic model and persists under disorder.
Second, within each band gap, isolated points appear either in pairs or singly, depending on $\kappa$ and $N$. Analysis of the corresponding wave functions confirms that these isolated points represent topological edge states. Importantly, the positioning of the nodes is dictated by the nontrivial Zak phase of the bulk bands, consistent with bulk–boundary correspondence: a Zak phase of
$\pi$ enforces edge-localized modes whose amplitudes vanish on specific sublattice sites adjacent to the short bonds, while a Zak phase of 0 yields only extended bulk states. Chiral symmetry further enforces sublattice polarization, explaining the vanishing amplitudes and ensuring that the nodal structures are symmetry-protected. Consequently, these edge states occur precisely at the discrete energies given by Eq.~(\ref{eq:eigen}), underscoring their topological protection within the disordered mosaic lattice. In contrast, localized bulk states are distributed arbitrarily throughout the system.

In line with the above discussion, all topological edge states in the disordered off-diagonal mosaic lattice exhibit intrinsic nodal patterns. This is illustrated in the inset of Fig.~\ref{ff7}(b) for $\kappa=3$ and Fig.~\ref{ff8}(a) and (c) for $\kappa=2$ and 4, where the wave functions vanish at every site $j\kappa$ for any integer $j$.
By contrast, in the diagonal disordered mosaic lattice, recent studies have identified discrete energy values [Eq.~(\ref{eq:eigen})] at which eigenstates display anomalous power-law localization \cite{Ngu1, Ngu2}. These states likewise exhibit nodal structures pinned to the mosaic-modulated sites, but their origin lies in resonant destructive interference at special energies rather than in a topological invariant. Consequently, they lack bulk–boundary correspondence and symmetry protection, making them fragile to perturbations.

To test the robustness of these topological states under different disorder realizations, Fig.~\ref{ff9} shows the IPR as a function of $E$ for five independent random configurations in a chain of size $L = 2001$ with $\kappa=3$ and $\beta_0=1.5$. Each band gap contains ten isolated points corresponding to the ten edge states predicted for these configurations, whose eigenvalues remain pinned exactly at $\pm 1$, demonstrating their robustness against disorder. Similar robustness is observed for the edge-state eigenvalues at $E=0$ for $\kappa=2$ and at $E=0,~\pm\sqrt{2}$ for $\kappa=4$ in Figs.~\ref{ff7} and \ref{ff8}. This invariance arises from the chiral symmetry of the off-diagonal mosaic lattice, which enforces spectral symmetry about zero (shifted by $\epsilon$). As discussed in Sec.~\ref{sec:level22}, the decoupling of the recursion relations ensures that the pinned values are protected by this symmetry. Thus, the robustness of the edge-state eigenvalues is an exact consequence of the lattice structure and its symmetry, not a statistical effect of disorder averaging.

Figure~\ref{ff10} presents results for a single random realization ($L=2001$, $\kappa=3$, $\beta_0=2$) at three different disorder strengths $W$. As expected in disordered systems, both edge and bulk states become more localized with increasing $W$, as indicated by larger IPR values, while the band gaps gradually shrink. Notably, the edge-state eigenvalues remain strictly pinned at $E=\pm 1$.

Finally, in Fig.~\ref{ff11}, we examine the disorder dependence of the $\beta$-dependent edge states for $\kappa=4$ and $\beta_0=1.5$.
For
$W\le 0.2$, these states consistently appear in all random realizations. As
$W$ increases slightly above 0.2, their presence becomes highly configuration-dependent: they may appear in some realizations but not in others, and their eigenvalues can even shift outside the gaps. Defining the critical disorder strength
$W_c$ as the value at which these states occur in roughly 50\% of realizations, we obtain
$W_c\sim 0.21$ for
$\beta_0=1.5$. We also find that $W_c$ increases gradually with increasing
$\beta_0$. This extremely small critical value underscores their fragility and stands in sharp contrast to the robustness of topological edge states, which remain intact even at
$W=2$ (Fig.~\ref{ff10}).

\section{\label{sec:level1} Conclusion}

We have systematically analyzed the emergence and robustness of topological edge states in 1D off-diagonal mosaic lattices with periodic and disordered hopping modulations. By extending the SSH model to arbitrary mosaic periods
$\kappa\ge 2$, we derived exact analytical expressions for the edge-state eigenvalues,
$E=\epsilon+2t\cos(\pi i/\kappa)$, and established that topological edge states are not confined to a few special cases, but are a generic feature of multi-band mosaic lattices. Numerical simulations confirmed these predictions, revealing characteristic nodal structures whose form depends sensitively on the arrangement of long and short bonds at the lattice boundaries.

We further examined the effects of off-diagonal disorder and found that topological edge states retain both their localization and eigenvalues even under substantial hopping fluctuations, underscoring their robustness. By contrast, the additional
$\beta$-dependent edge states that arise for
$\kappa\ge 4$
are fragile and disappear as the disorder strength increases.

Overall, our results provide a unified framework for understanding and engineering multi-gap topological phases in off-diagonal systems. They open avenues for realizing disorder-resilient edge states in synthetic quantum materials, photonic crystals, and other engineered lattices. Future directions include exploring their dynamical properties, responses to external fields, and potential applications in robust waveguiding and topological quantum information processing.

\section*{Acknowledgments}
This work was supported by the National Research Foundation of Korea (NRF) grant funded by the Korean Government (RS-2025-16071339). It was also supported by the Basic Science Research Program through the NRF, funded by the Ministry of Education (RS-2021-NR060141).

\appendix

\section{Derivation of $\beta$-dependent edge states for $\kappa=4-6$}
Let the left edge start with a short bond of strength
$\beta$ (on-site energies $=0$). The boundary equations are
\begin{align}
E\psi_1&=\beta\psi_2,\nonumber\\
E\psi_2&=\beta\psi_1 +\psi_3,\nonumber\\
E\psi_3&=\psi_2 +\psi_4,\nonumber\\
E\psi_4&=\psi_3 +\psi_5,\nonumber\\
E\psi_5&=\psi_4 +\beta\psi_6.
\label{eq:bes}
\end{align}
Assume an edge state with
$\psi_{i+\kappa}=r\psi_i$ and $\vert r\vert<1$.
For $\kappa=4$, this gives $\psi_5=r\psi_1$ and $\psi_6=r\psi_2=(rE/\beta)\psi_1$. The last line gives
\begin{align}
    Er\psi_1=\psi_4+rE\psi_1 ~\Rightarrow~\psi_4=0.
\end{align}
Using the first three lines with $\psi_4=0$ yields
\begin{align}
    E(E^2-\beta^2-1)=0~\Rightarrow ~E=\pm\sqrt{1+\beta^2},
\end{align}
discarding $E=0$ (non-normalizable). The decay factor is
\begin{align}
r=-\frac{E}{\beta^2}=\mp\frac{\sqrt{\beta^2+1}}{\beta^2},
\end{align}
so $\vert r\vert<1$ when $\beta>1$.
For larger $\kappa$, the same procedure gives the $\beta$-dependent Tamm-like branches. In particular,
\begin{align}
    \kappa=5:~~E=&\pm\sqrt{\frac{\beta^2+2\pm\sqrt{\beta^4+4}}{2}},\nonumber\\
    \kappa=6: ~~E=&\pm\sqrt{\frac{\beta^2+3\pm\sqrt{\beta^4-2\beta^2+5}}{2}}.
\end{align}

\end{document}